\documentclass[twocolumn]{aa}
\usepackage{amsmath}
\usepackage{caption}
\usepackage{subcaption}
\usepackage{txfonts}
\usepackage{xcolor}
\usepackage{hyperref}
\hypersetup{
    colorlinks=true,
    allcolors=blue
}

\graphicspath{{images/}}

\newcommand{\angstrom}{\textup{\AA}}
\newcommand{\Rsun}{{$\mathrm{R}_{\odot}$}}
\newcommand{\kmps}{{$\mathrm{~km~s}^{-1}$}}
\newcommand{\mpss}{{$\mathrm{~m~s}^{-2}$}}

\begin{document}

\title{On the three-dimensional relation between the coronal dimming, erupting filament and CME}
\subtitle{Case study of the 28 October 2021 X1.0 event}

\titlerunning{Coronal dimming - a proxy for the directivity of filament/CME}
\authorrunning{Chikunova et al.}

\author{Galina Chikunova\inst{1}
   \and Tatiana Podladchikova\inst{1}
   \and Karin Dissauer \inst{2}
   \and Astrid M. Veronig\inst{3,4}
   \and Mateja Dumbovi\'c\inst{5}
   \and Manuela Temmer\inst{3}
   \and Ewan C.M. Dickson\inst{3}}

\institute{Skolkovo Institute of Science and Technology, Bolshoy Boulevard 30, bld. 1, 121205, Moscow, Russia \\
\email{Galina.Chikunova@skoltech.ru}
\and 
NorthWest Research Associates, 3380 Mitchell Lane, Boulder 80301, CO, USA
\and
University of Graz, Institute of Physics, Universit\"atsplatz 5, 8010 Graz, Austria
\and
University of Graz, Kanzelh\"ohe Observatory for Solar and Environmental Research, Kanzelh\"ohe 19, 9521 Treffen, Austria
\and
Hvar Observatory, Faculty of Geodesy, University of Zagreb, Kaciceva 26, HR-10000, Zagreb, Croatia
}

\date{Received May, 2023; accepted August, 2023}

\abstract
{Coronal dimmings are localized regions of reduced emission in the extreme-ultraviolet (EUV) and soft X-rays, formed due to the expansion and mass loss by coronal mass ejections (CMEs) low in the corona. Distinct relations have been established between coronal dimmings (intensity, area, magnetic flux) and key characteristics (mass, speed) of the associated CMEs by combining coronal and coronagraphic observations from different viewpoints in the heliosphere.
}
{We investigate the relation between the spatiotemporal evolution of the dimming region and the dominant direction of the filament eruption and CME propagation for the 28 October 2021 X1.0 flare/CME event observed from multiple viewpoints in the heliosphere by Solar Orbiter, STEREO-A, SDO, and SOHO.
}
{We present a method to estimate the dominant direction of the dimming development based on the evolution of the dimming area, taking into account the importance of correcting the dimming area estimation by calculating the surface area of a sphere for each pixel. To determine the flux rope propagation direction during the early CME evolution, we perform 3D reconstructions of the white-light CME by graduated cylindrical shell modeling (GCS) and 3D  tie-pointing of the eruptive filament.
}
{The dimming evolution starts with a radial expansion and later propagates more to the southeast. The orthogonal projections of the reconstructed height evolution of the prominent leg of the erupting filament onto the solar surface are located in the sector of the dominant dimming growth, while the orthogonal projections of the inner part of GCS reconstruction align with the total dimming area. The filament reaches a maximum speed of $\approx$250\kmps~at height of about $\approx$180~Mm before it can no longer be reliably followed in the EUV images. Its direction of motion is strongly inclined from the radial direction (64$^\circ$ to the East, 32$^\circ$ to the South). The 3D direction of the CME and the motion of the filament leg differ by 50$^\circ$. This angle roughly aligns with the CME half-width obtained from the CME reconstruction, suggesting a relation of the reconstructed filament to the associated leg of the CME body. 
}
{The dominant propagation of the dimming growth reflects the direction of the erupting magnetic structure (filament) low in the solar atmosphere, though the filament evolution is not related directly to the direction of the global CME expansion. At the same time, the overall dimming morphology closely resembles the inner part of the CME reconstruction, validating the use of dimming observations to obtain insight into the CME direction.
}


\maketitle

\section{Introduction} \label{sec:intro}

Coronal Mass Ejections (CMEs) are large-scale eruptions in the solar corona, being associated with huge energy release \citep{harrison1995nature}. These are huge quantities of the magnetized plasma with masses up to 10$^{13}$ kg that are ejected from the Sun into interplanetary space with speeds in the range of $\approx$100-3000 \kmps~\citep{2009EM&P..104..295G, webb2012coronal}. Earth-directed events can lead to a multitude of space weather effects in geospace, including power networks knockout, disturbances of the radio signal, and various anomalies of satellites \citep[e.g. reviews by][]{2007LRSP....4....1P,2017SSRv..212.1137K,temmer2021}. 
However, the measurements of CME parameters, especially for Earth-directed events, observed as halo-type events for spacecraft located along the Sun-Earth line, may be strongly affected by the projection effects \citep{Burkepile2004,gopalswamy2000interplanetary}.
Moreover, traditional instruments to analyze them, such as coronagraphs, occult the early part of the CME evolution in the low corona. The analysis of CME manifestations in the low corona, including filament eruptions and coronal dimmings, is thus important in understanding the initiation mechanism and the early evolution of CMEs.

Coronal dimmings are regions of strongly reduced emission in extreme-ultraviolet (EUV) \citep{Thompson1998,Zarro1999} and soft X-ray \citep{Hudson:1996,Sterling:1997} wavelengths, occurring in association with CMEs. They are direct signatures of the expansion and ejection of coronal plasma by the CME \citep{Thompson1998, Harrison&Lyons2000}. The mass loss nature of their appearance is supported by the presence of plasma outflows detected in spectroscopic observations in the EUV \citep[e.g.][]{harrison2000spectroscopic,harra2001material,Tian2012, Veronig2019} and the intensity decrease observed in multiple wavelengths \citep{Zarro1999, chertok2003large}. Moreover, Differential Emission Measure (DEM) diagnostics reveals impulsive drops in coronal dimming regions by up to $70\%$ of the pre-event values \citep{lopez2017mass,2018ApJ...857...62V,Veronig2019}. Recent statistical studies of coronal dimming events observed on-disk \citep{2016ApJ...830...20M,2016ApJ...831..105A, 2017ApJ...839...50K, dissauer2019statistics} by Solar Dynamics Observatory \citep[SDO;][]{2012SoPh..275....3P} and off-limb \citep{chikunova2020} by Solar TErrestrial Relations Observatory \citep[STEREO;][]{Kaiser:2008} satellites reveal distinct correlations of characteristic dimming parameters with the speed and mass of the associated CMEs.

During the history of their observations, coronal dimmings were widely acknowledged as reliable indicators of front-side CMEs \citep{Thompson2000}, as they appear on the visible part of the solar disk. More detailed monitoring \citep{mandrini2007cme,2009SPD....40.1708W,Temmer2011,dissauer2018statistics} and magnetohydrodynamical (MHD) simulations \citep{downs2012understanding, prasad2020magnetohydrodynamic} show that the locations of the so-called ``core'' dimmings map the footpoints of the erupting flux rope, while the spatial growth of the more wide-spread secondary dimmings indicates the expansion of the overlying magnetic structures of the CME expansion in the low corona. 
Together with other CME manifestations in the corona, dimmings reveal crucial information about the early configuration of the eruption and its initial parameters \citep[e.g. case studies by][]{moestl2015, temmer2017flare, dissauer2018detection, heinemann2019, dumbovic_Womenday, thalmann2022}.
The importance of coronal dimmings for the study of solar CME/flares, as well as the latest significant results in this area, are discussed in \cite{2022SoPh..297...59K}. The authors of the SunCET mission concept \citep{2021JSWSC..11...20M} suggest using coronal dimmings as a source of information on Earth-directed halo CME kinematics, observed in EUV rather than visible light. \cite{veronig2021indications} proposed a new approach of using coronal dimmings as indicators of CMEs on solar-type stars, expanding their potential for studying stellar physics.

In this study, we investigate an important aspect of the dimming-CME relationship: the evolution of the dimming area (projected onto the 2D image plane) and how it relates to the directivity of the different parts of the erupting magnetic structure, namely the filament and the CME (both reconstructed in 3D using multi-viewpoint observations).
In the past years, numerous studies have focused on the 3D reconstruction of CMEs using dual/triple coronagraphic observations from the STEREO and SOHO spacecraft. 
The tie-pointing method \citep{mierla2008,byrne2010propagation, Liewer2011,liu2010reconstructing} identifies the same feature in images taken from different viewpoints and use of triangulation to determine its position in 3D \citep{thompson2006coordinate}. Use of the epipolar geometry \citep[introduced by][]{Inhester2006} helps to better constrain the tie-points to a given straight line and reduces the matching problem. Another approach involves the reconstruction of an associated filament, which is located within the CME flux rope and contains the material of the CME bright core \citep{Illing1986,li2013fine}. \cite{Liewer2009} used stereoscopic tie-pointing for the 19 May 2007 erupting filament reconstruction and compared the filament appearance prior to and after the eruption to reveal the place of magnetic reconnection. \citep{2014ApJ...790...25S} combined a 3D-reconstruction of the erupting filament with a polarization ratio technique \citep{2004Sci...305...66M}, for a better determination of the CME kinematics. To examine the 3D development of CMEs resulting from non-radial filament eruptions, \cite{zhang2021revised} used a cone model, placing an apex of the cone at the source region of an eruption instead of the Sun center as in the traditional cone models. A different approach is forward modeling from coronagraphic observations.  Graduated Cylindrical Shell (GCS) modeling \citep{2006ApJ...652..763T,thernisien2011implementation} of CMEs uses a strongly idealized flux rope structure, composed of two cones, presenting the legs of a CME, and a torus-like structure connecting these cones, so-called a “hollow croissant” shape. \cite{2010ApJ...722.1762L} compared the GCS modeling with the flux rope reconstruction using in-situ measurements at 1 AU and showed that model can be used to obtain estimations of CME position, direction, three-dimensional extent, and speed. These and other methods of reconstructing the CME direction are described and compared in detail in \cite{mierla20103}. All of them require the presence of at least two simultaneous observations of the CME from different vantage points and significantly depend on the presence of the STEREO satellite. 

We look closer at the source region of an event and try to understand if the single-point observations of the coronal dimmings can indicate the initial direction of the erupting filament and/or CME propagation. Coronal dimmings are very well observed by different EUV instruments, providing unique information about the pre-CME evolution. Nowadays we have regular high-cadence high-resolution imaging of the solar EUV corona by the Atmospheric Imaging Assembly \citep[AIA;][]{2012SoPh..275...17L} onboard SDO, the Extreme Ultraviolet Imager \citep[EUI;][]{2020A&A...642A...8R} onboard Solar Orbiter \citep[SolO;][]{muller20}, the Solar Ultraviolet Imager \citep[SUVI;][]{darnel2022goes} onboard the Geostationary Operational Environmental Satellites (GOES), as well as the STEREO-A Extreme-Ultraviolet Imaging Telescope  \citep[EUVI;][]{Wuelser2004}. These observations provide us with various possibilities to monitor the CME low coronal signatures, increasing the chances to detect the potential geoeffective eruptions at a very early stage, i.e. even before their front reaches the FOV of coronagraphs.

In this study, we use EUV observations of the Sun and develop a new method to track coronal dimming parameters in terms of direction and temporal evolution, and compare them with the evolution of the eruptive filament and CME for the 28 October 2021 event which was related to an X1.0 flare. By projecting the reconstructed filament heights on the solar surface, we investigate the dominant propagation of the dimming evolution and study its relation to the main propagation direction of the erupting filament and the CME.

\section{Data and Event Observation} \label{sec:data}
The event under study took place on 28 October 2021 from NOAA active region (AR) 12887. The CME and filament eruption were associated with an X1.0 flare (start: 15:17 UT, peak: 15:35 UT) at heliographic position S28W01. The event also produced a globally propagating EUV wave and a Ground level enhancement (GLE), which were studied independently by different research groups \citep{devi2022extreme,hou2022three,2022A&A...660L...5P,2022A&A...663A.173K}. The three-dimensional velocity of the  plasma outflows associated with the event was estimated in Sun-as-a-star spectroscopic observations in \cite{sunasstar2022} resulting in $v\approx 600$~\kmps. A data-driven simulation of the eruption was presented in \cite{guo2023thermodynamic}. The authors discuss that they found good agreement for including the morphology of the eruption, the kinematics of the flare ribbons, extreme ultraviolet (EUV) radiations, and the two components of the EUV waves predicted by the magnetic stretching model.

\begin{figure}
\centering
\includegraphics[width=0.85\columnwidth]{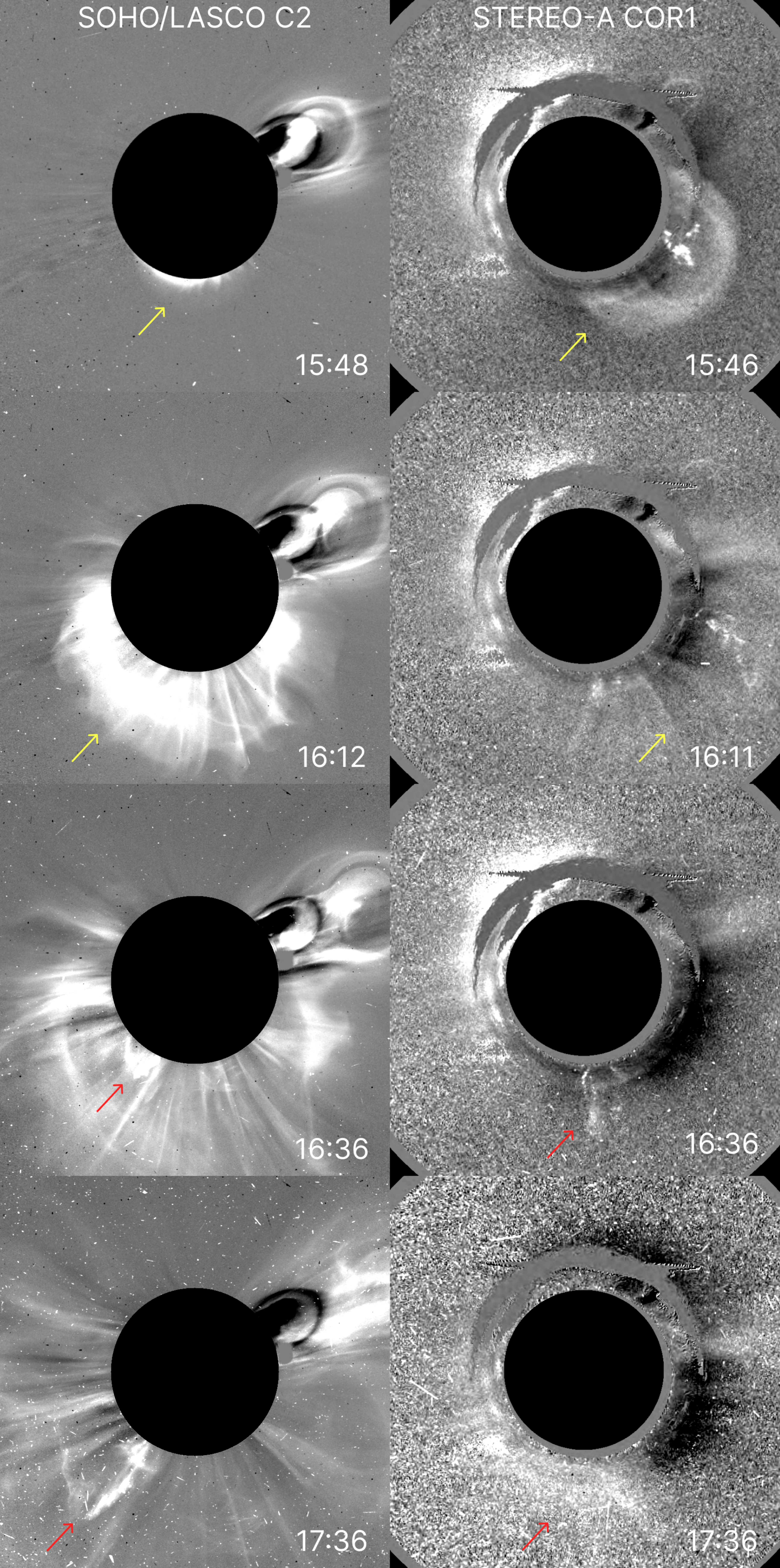}
\caption{Snapshots of coronagraphic observations, using JHelioviewer \citep{JHelioviewer}. Left column: SOHO/LASCO C2 white-light base difference images. Right column: white-light base difference images from STEREO-A COR1. Yellow arrows indicate the CME and red arrows show the propagated leg of the filament eruption at time steps 16:36~UT and 17:36~UT, respectively. 
}
\label{fig:coronagraphs}
\end{figure}

The CME appeared as a halo in the SOHO/LASCO C2 field of view (FOV) starting at 15:48 UT. Figure~\ref{fig:coronagraphs} shows simultaneous observations of its evolution by SOHO/LASCO~C2 (left panels) and STEREO-A EUVI COR2 coronagraphs (right panels). The brightest part of the CME (yellow arrows) propagates southward for LASCO, while STEREO/COR1 shows the signatures of a three-part CME structure \citep{Illing1986, Low1995, Vourlidas2013, Cheng2017, Howard2017, Veronig2018}. Around 16:36~UT we observe a second propagating part, a flank of the CME (red arrows).
The LASCO catalog\footnote{\url{https://cdaw.gsfc.nasa.gov/CME\_list/}} reports a plane-of-sky speed of the CME of about 1500~\kmps.

\begin{figure}
\includegraphics[width=1\columnwidth]{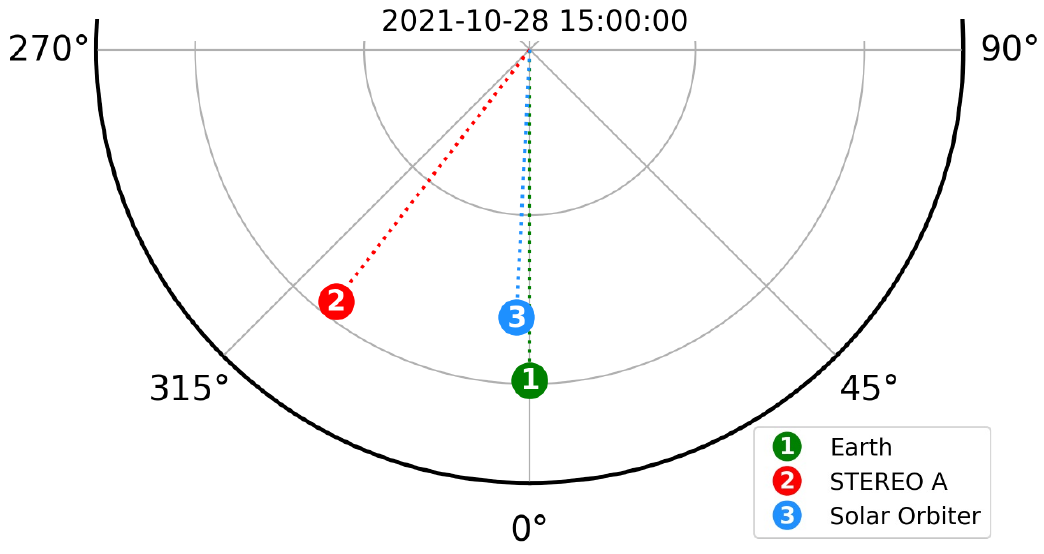}
\centering
\caption{A view of the ecliptic plane showing the various spacecraft positions on 28 October 2021. 
SDO is located along the Sun-Earth line, while the STEREO-A and Solar Orbiter spacecraft are positioned $37.6^{\circ}$ and $2.8^{\circ}$ to the East, respectively. 
The image is generated using the Solar-MACH tool \citep{SolarMach2023}.}
\label{fig:satellites}
\end{figure}

The EUV low coronal signatures of the CME were simultaneously observed by the SDO/AIA and STEREO-A/EUVI, separated by $37.6^{\circ}$ in longitude at this date (see Figure~\ref{fig:satellites}). The AIA instrument provides images of the Sun in seven EUV filters with a pixel scale of 0.\arcsec 6 over a
FOV of about 1.3\Rsun. EUVI observes the Sun through four EUV passbands over a FOV up to 1.7\Rsun~providing images with a pixel scale of 1.\arcsec 6. 

For the analysis of coronal dimmings we used AIA 193~\angstrom~filtergrams images with a 12s cadence. The selection of 193~\angstrom~for observing and extracting coronal dimmings is motivated by the systematic study of \cite{dissauer2018detection}, demonstrating that this wavelength effectively captures both the quite and AR coronal plasma that is expelled during CMEs. The same images effectively resolved the filament propagation and we matched them with 2.5~min cadence sequences of 195~\angstrom~EUVI data for a 3D reconstruction of the filament. Both data series were calibrated and corrected for differential rotation to a reference time of 28~October~15:05~UT, using the Sunpy library \citep{sunpy_community2020} in Python. The detection and analyzing procedures are also developed using Python.

\begin{figure}
\centering
\includegraphics[width=1\columnwidth]{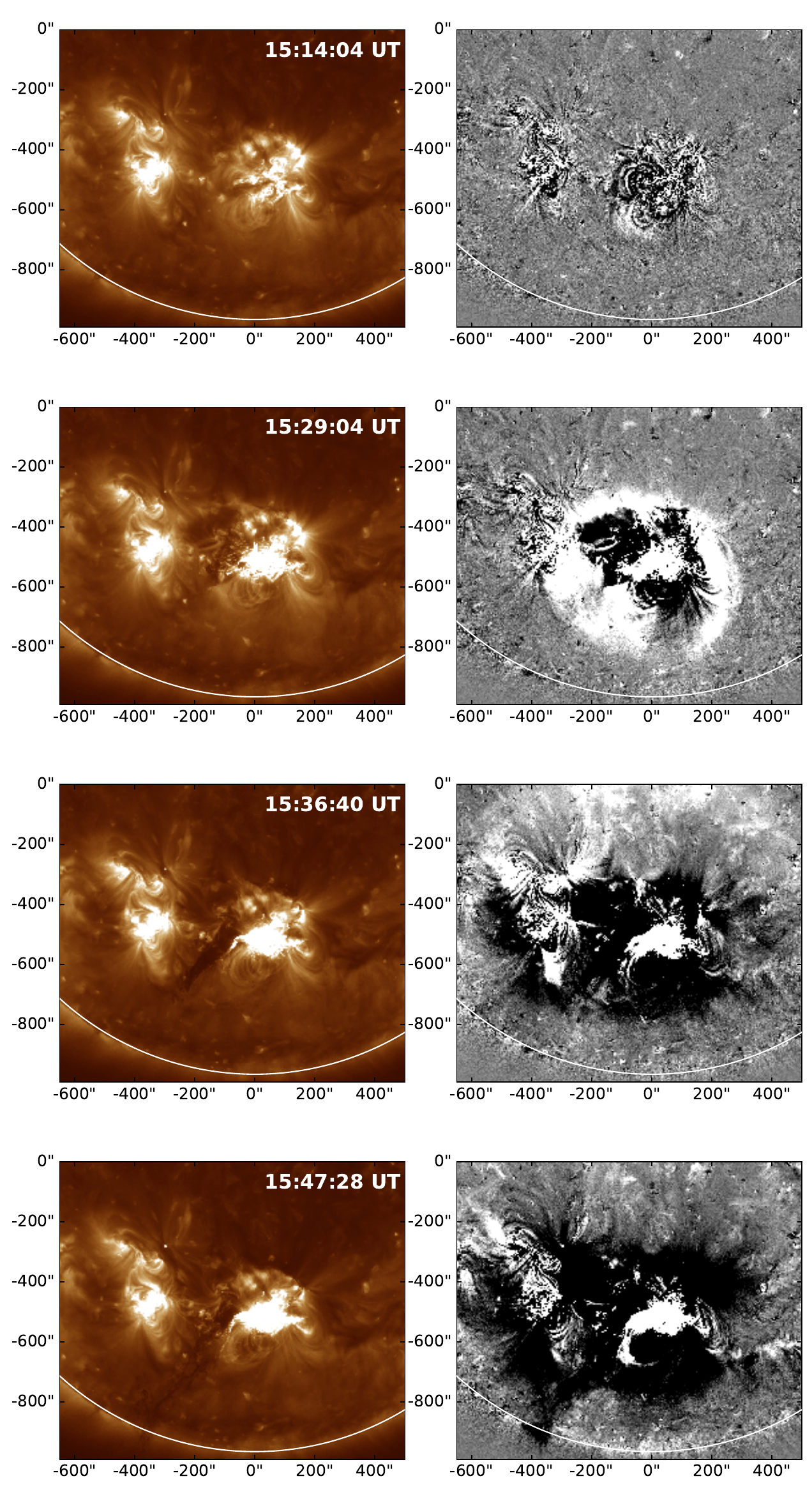}
\caption{Overview of the 28 October 2021 event, showing the dimming evolution and eruption of the filament in AIA 193~\angstrom. Left: direct images; right: corresponding base-difference images. Four time steps are shown, while the full evolution of the event can be found in the accompanying movie.}
\label{fig:event_overview}
\end{figure}

Figure~\ref{fig:event_overview} gives an overview of the event as observed by the AIA original 193~\angstrom~filter showing direct images (left) and processed base-difference images (right). 
The coronal dimming, eruptive filament, and EUV wave are well observed in base difference data, originating at longitudes close to the disk center for the line of sight (LOS) from Earth. The dimming is detected from the earliest time steps, starting around 15:05~UT with quasi-circular propagation, which later dominates mostly to the South-East, reaching the neighboring active region. The filament eruption starts around 15:10-15:12~UT, propagating to the South-East and reaching off-limb at 15:47~UT. Another part of the filament is visible off-limb around 15:35 UT to the West. The EUV wave appears around 15:28~UT and is visible on the solar surface up to 15:50~UT, showing an initial quasi-circular propagation, with dense fronts towards the north direction (second and third panels, see also the accompanying movie).  \cite{devi2022extreme} presented a full analysis of the EUV wave evolution, reporting the maximum speed of its propagation of about 700 ~\kmps.

\begin{figure}
\centering
\includegraphics[width=1\columnwidth]{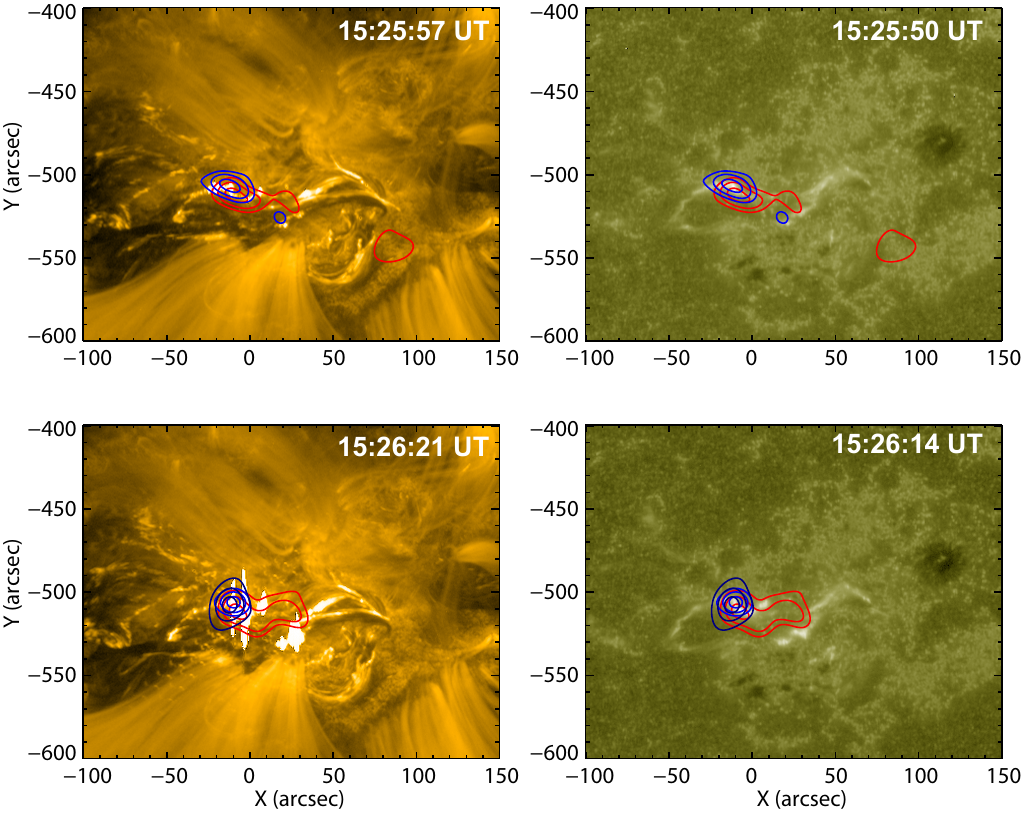}
\caption{STIX images showing 50, 70 and 90\% contours for energy bands 4 -- 10~keV (red), 15 -- 25~keV (blue) and 25 -- 50~keV (dark blue). The STIX contours are rotated to SDO's point of view and overlaid on AIA images.  
The AIA images are shown for two different time steps -- top row:  15:25~UT and bottom row: 15:26~UT; 
and at two different AIA filters -- left column: 171~\angstrom~and right column: 1600~\angstrom.
}
\label{fig:stix_image}
\end{figure}

The associated X1.0 flare was also detected in hard X-rays (HXR) by the Spectrometer Telescope for Imaging X-rays (STIX) \citep{2020A&A...642A..15K} on board Solar Orbiter. On this date Solar Orbiter was close to the Sun-Earth line ($2.8^{\circ}$ East) at a distance of 0.80 AU from the Sun.
As this was an X-class flare, STIX detected significant emission in both the thermal and non-thermal energy ranges, enabling insight into a broader range of the flare process. 
STIX uses a pair of grids to indirectly reconstruct images of the HXR photons by producing Moire fringes that deferentially illuminate different detector pixels. STIX images (Figure~\ref{fig:stix_image}) were produced in three energy bands: the thermally dominated 4--10~keV, the intermediate 15--25~keV, and the non-thermal  25--50~keV band. 
For context and comparison, these were overlaid on SDO AIA 171~\angstrom~and 1600~\angstrom~images after being rotated to the SDO point of view. In order to accurately perform the rotation of the STIX sources, estimates of the X-ray source heights must be made. For the non-thermal footpoints the equation in \citet{2002SoPh..210..383A} relating energy to height was used. For the thermal sources a semi-circle connecting the two ribbons was assumed. As there was no STIX aspect solution for this time a small adjustment of [$-40$\arcsec, 5\arcsec] was applied to align the STIX sources with the AIA ribbons. The STIX observations were corrected for the difference in light travel time (94.58~s) from the Sun to Solar Orbiter and from the Sun to Earth.

\section{Methods and Results} 
Our approach consists of three steps. First, we develop a method of estimating the dominant direction of the dimming development (Section~\ref{sec:dimmings}). Second, to determine the direction of parts of the erupting structure/flux rope during the early CME evolution, we perform 3D reconstructions of the eruptive filament (Section~\ref{sec:filament}) and a 3D CME reconstruction using GCS modeling (Section~\ref{sec:GCS}). Finally, we study the relation between the dominant dimming direction, the direction of the filament eruption, and the CME (Section~\ref{sec:direction}).

\subsection{Coronal Dimming} \label{sec:dimmings}
Here, we present the method to estimate the dominant direction of the dimming development, which includes the dimming segmentation (Section~\ref{sec:dimming_segment}), followed by the assessment of the evolution of the dimming area with the help of sector analysis (Section~\ref{sec:dimming_sector}).

\subsubsection{Dimming segmentation}\label{sec:dimming_segment}
For the dimming segmentation and analysis, we use a sequence of AIA $193~\angstrom~$maps within the 15:00-15:57~UT time range. As the dimming regions are characterized by a drop in EUV intensity, for their detection we create base-difference images, indicating the absolute change in the intensity values with respect to the pre-eruptive coronal state. An automated detection technique, based on thresholding and region growing is applied to these images (similar to the one described in \cite{dissauer2018detection}). Firstly, we extract all pixels, where the intensity dropped below $-30$~DN. This level was selected empirically by qualitatively checking images during different development stages of the dimming in order to include also distant parts (see Figure~\ref{fig:event_overview}). Secondly, we use 30\% of the darkest pixels of this set as seeds for the region-growing algorithm to connect all the separated parts of the dimming region. To minimize the noise we use median filtering with a $3\times3$ kernel.

We obtain the dimming masks for all the time steps from 15:05 to 15:57~UT and unite them into one cumulative dimming mask. This mask contains all pixels that have been flagged as dimming pixels during the chosen time range. However, pixels of the erupting filament also show up in the mask as they have a similar intensity level. To exclude these pixels we use the same mask in a base-difference image at one of the last time steps (when the filament was already gone) and refine the mask by deleting all the pixels with intensity above $-30$~DN, subsequently filling the gaps within the contour. The final cumulative dimming mask (\ref{fig:dimming_bd_time}) is then used for further dimming analysis. 

\subsubsection{Dimming sector analysis}\label{sec:dimming_sector}
To determine the main direction of the dimming expansion, we design a spherical polar coordinate system centered at the source region and divide the solar surface into 16 angular sectors of $\Delta\phi=22.5^{\circ}$ width (counted counterclockwise). The source location was defined in accordance with the center of the flare emission contours from STIX (Figure \ref{fig:stix_image}).
For each sector, we further calculate the integral area of all pixels by developing a method to estimate the surface area of a sphere for every pixel. In appendix~\ref{sec:appendix}, we present a comprehensive description of the area calculation, including a comparative analysis of sector areas derived from corrected pixels and their corresponding averages. 
we made all the code publicly available on GitHub\footnote{\url{https://github.com/Chigaga/area_calculation}}.

\begin{figure*}
\centering
\includegraphics[width=2\columnwidth]{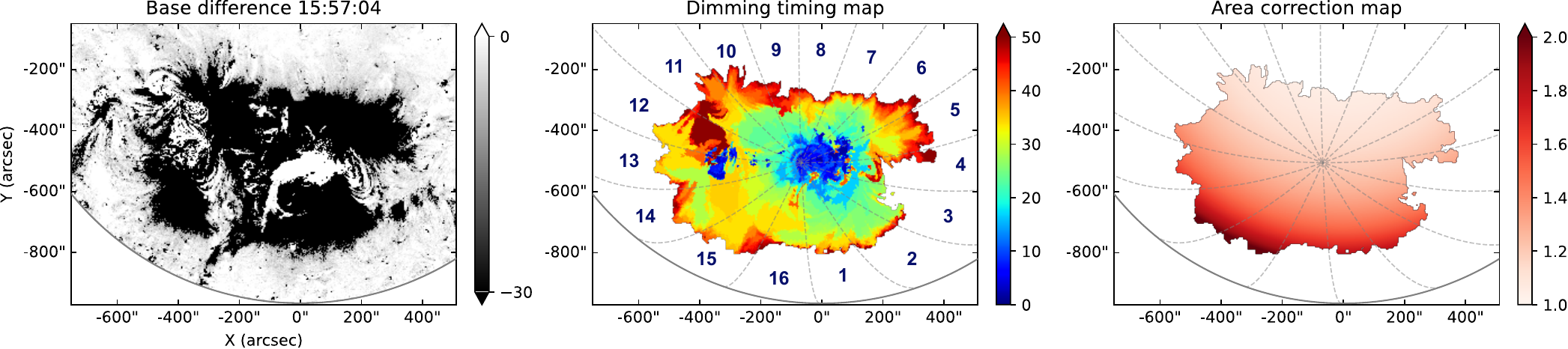}
\caption{Total area of the dimming region. Left: the base-difference map at the time of maximal dimming extent at 15:57~UT. Middle: cumulative timing map of the whole dimming extraction, color-coded in minutes from 15:05~UT. Right: the same dimming map, but each pixel shows the ratio of its area to the area of the average pixel in the center. Dashed curves and numbers represent the chosen sectors.} 
\label{fig:dimming_bd_time}
\end{figure*}

Figure~\ref{fig:dimming_bd_time} (left panel) shows the dimming region at the times close to its maximal extent \citep[similar to the end of the dimming impulsive phase described in][]{dissauer2018statistics}, the cumulative dimming map, where each pixel is color-coded by the time of its first detection (middle panel) and by the ratio of each corrected pixel area to the average (square) pixel area in the center of the Sun (right panel) for each sector. 

\begin{figure}
\centering
\includegraphics[width=1\columnwidth]{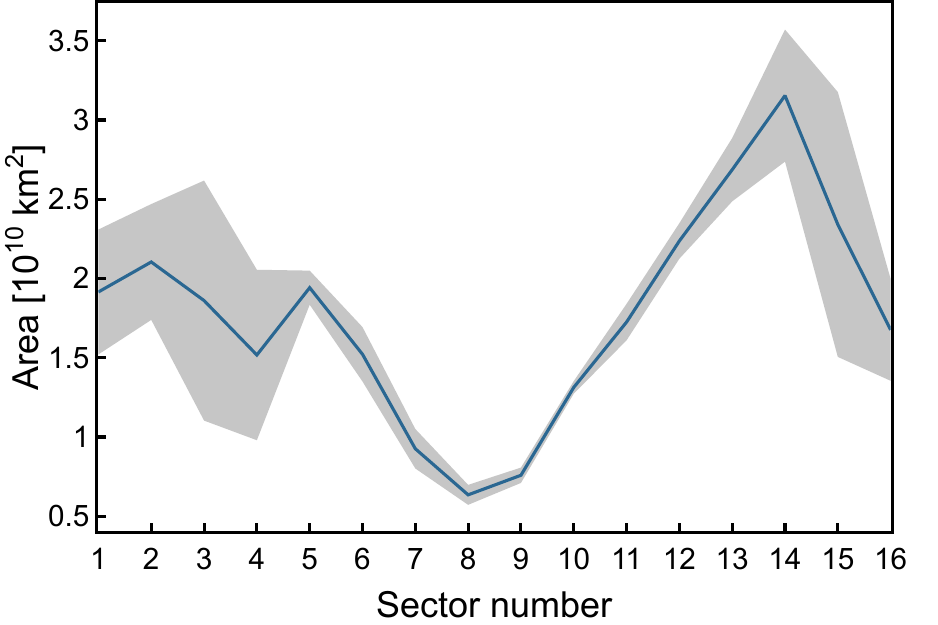}
\caption{Dimming area calculated for each sector in the full cumulative map. The shaded band indicates the error estimation using different threshold levels used for the dimming segmentation. The largest dimming area is in sector no.14.}
\label{fig:area}
\end{figure}

We define the middle longitude of the sector with the largest area as the main direction of the overall dimming development. To estimate the uncertainty in the area due to a specific threshold used for the dimming segmentation, we additionally apply a $\pm$5\% change to the threshold level and compute the mean values and corresponding standard deviation of the dimming area in each sector. The resulting dimming areas for the different sectors together with the uncertainty ranges are shown in Figure~\ref{fig:area}.

As can be seen, the dimming dominates in the southern sectors with the largest area in sector no. 14. Furthermore, analyzing the cumulative timing map (the middle panel in Figure~\ref{fig:dimming_bd_time}), we observe a two-stage evolution of the dimming region. Initially, the dimming spreads around the source, as indicated by the presence of blue and green pixels indicating evolution until $\approx$15:35~UT. Subsequently, the southeastern part of the dimming region becomes more pronounced and dominates the overall expansion.

In this particular case, the use of one cumulative mask in the dimming segmentation instead of tracking the dimming region over time helps to include the dimming areas hidden by other coronal structures at different stages of their evolution, e.g. initially covered by flare loops. At the same time, we note that the sector analysis can be also successfully implemented to a sequence of dimming masks, that allows to track the time evolution of characteristic dimming parameters (e.g. area/brightness/leading edge) during different parts of the dimming evolution with respect to the eruption source. The dominant direction of the dimming propagation can then be defined as the sector of the fastest dynamics, e.g. the highest area change rate for the particular time step. 

 \subsection{Eruptive Filament} \label{sec:filament}
Here, we present 3D reconstructions of the eruptive filament (Section~\ref{sec:3d_rec}) and study its kinematics  (Section~\ref{sec:filament_kinematics}).

 \subsubsection{3D-reconstruction of the erupting filament} \label{sec:3d_rec}
The dense and cool filament material is usually thought to be embedded in the dips of a magnetic flux rope, where the magnetic tension force keeps it in balance against the gravitational force. Therefore, the erupting filament can be interpreted to indicate the inner part of the erupting flux rope during the early stages of the eruption. The use of dual-point EUV observations from SDO and STEREO-A allows us to perform the 3D reconstruction of the eruptive filament using an epipolar geometry approach (see full description in \cite{Inhester2006, Podladchikova2019}). An object point and two observer positions create an epipolar plane, which, by definition, is seen as an epipolar line in each of the observer's images. At each time step, we manually identify the highest points of the filament from SDO/AIA $193~\angstrom$ images and match it with the corresponding points on the epipolar lines located in STEREO-A/EUVI $195~\angstrom$ data. Thus, for every point of the filament there is only one degree of freedom in placing the corresponding matching point in another image. In order to enhance the contrast of the erupting/moving features, we use base-difference images along with direct images for feature identification.

\begin{figure*}
\centering
\begin{subfigure}[t]{\linewidth}
\centering
\includegraphics[width=1\linewidth]{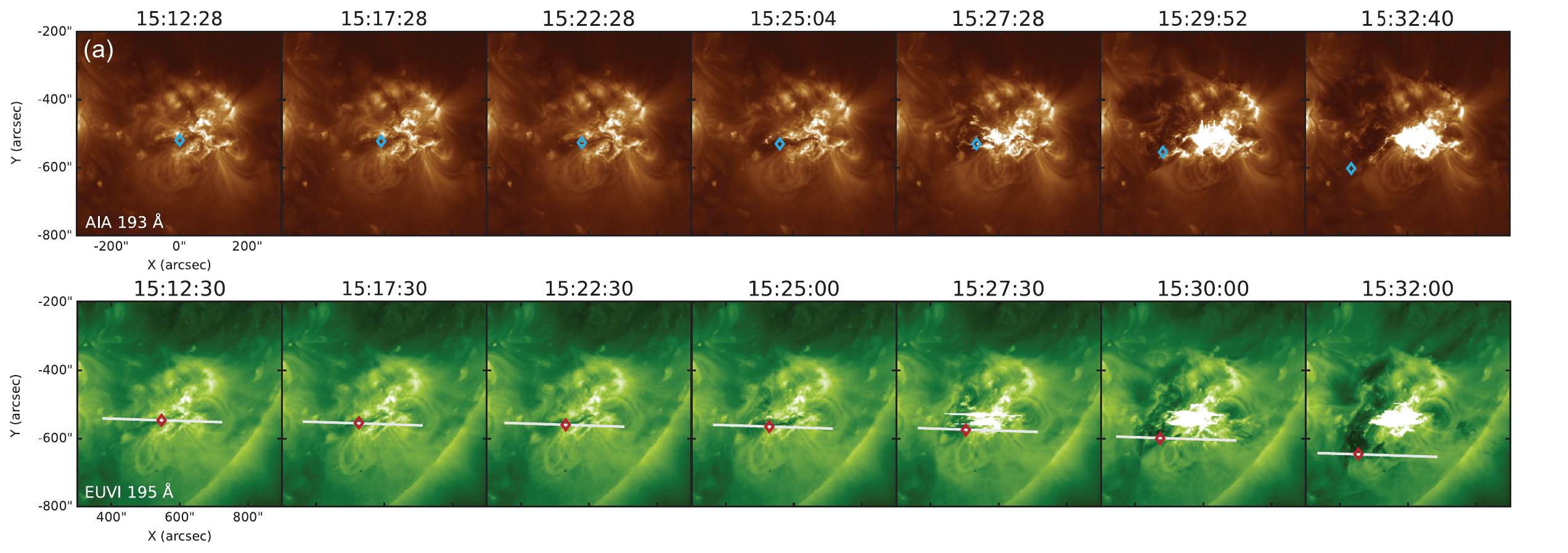} 
\label{fig:timing1}
\end{subfigure}
\begin{subfigure}[t]{\linewidth}
\centering
\includegraphics[width=1\linewidth]{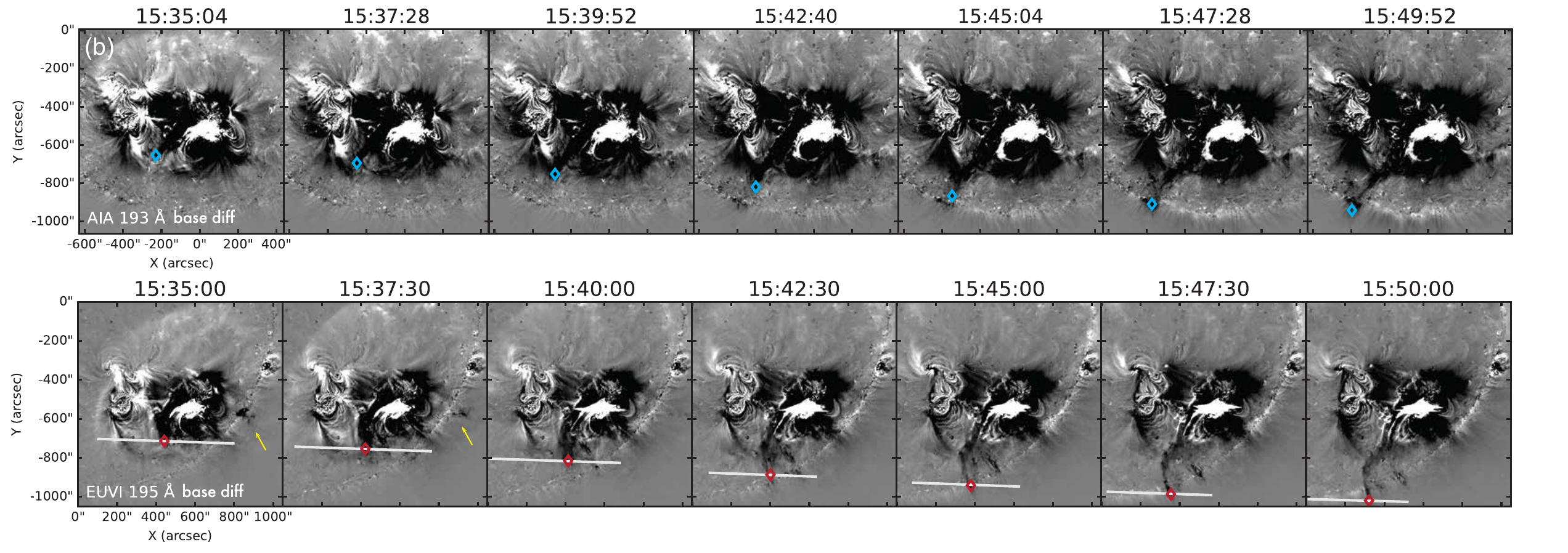} 
\label{fig:timing2}
\end{subfigure}
\caption{ Evolution of the erupting filament between 15:12--15:32~UT in SDO AIA $193~\angstrom$ (top) and STEREO-A EUVI $195~\angstrom$ direct images (a) and between 15:35--15:50~UT in base difference images (b). Blue markers show the upper tip of the filament, observed by SDO AIA, and red markers indicate the matching points from the STEREO-A vantage point. Epipolar lines are shown in white. The yellow arrows indicate another, central part of the filament eruption. Note, that the image scale of panels (a) and (b) is different. An animation of the reconstructions of panels (b) can be found in the accompanying movie.
}
\label{fig:3d_snapshots}
\end{figure*}

Figure~\ref{fig:3d_snapshots} shows a sequence of SDO/AIA $193~\angstrom$ (top) and STEREO-A/EUVI $195~\angstrom$ (bottom) maps, where the colored markers highlight the propagation of the erupting filament based on matching the same features on both AIA (blue) and EUVI (red) 
images. We track the filament growth using direct images for the time steps 15:05-15:32~UT (a) and base-difference images for 15:35-15:57~UT (b). For each point of the filament we obtain coordinates in the HEEQ coordinate system, defining its position in 3D. 

We then determine the orthogonal projections of the reconstructed 3D points on the solar surface, which reflect the projected direction of the filament eruption, using the following Equation 

\begin{align} 
X_P &=\frac{R_{Sun}\cdot X}{\sqrt{X^2+Y^2+Z^2}}\nonumber \\
Y_P &=\frac{R_{Sun}\cdot Y}{\sqrt{X^2+Y^2+Z^2}}  \\ 
Z_P &=\frac{R_{Sun}\cdot Z}{\sqrt{X^2+Y^2+Z^2}} \nonumber 
\end{align}

Here, $X_P$, $Y_P$, $Z_P$ denote the 3D cartesian coordinates of the orthogonal projection on the solar surface, $R_{Sun}$~is the solar radius, $X$, $Y$, $Z$ are the 3D coordinates of the reconstructed points of the filament. 

\begin{figure}
\centering
\includegraphics[width=1\columnwidth]{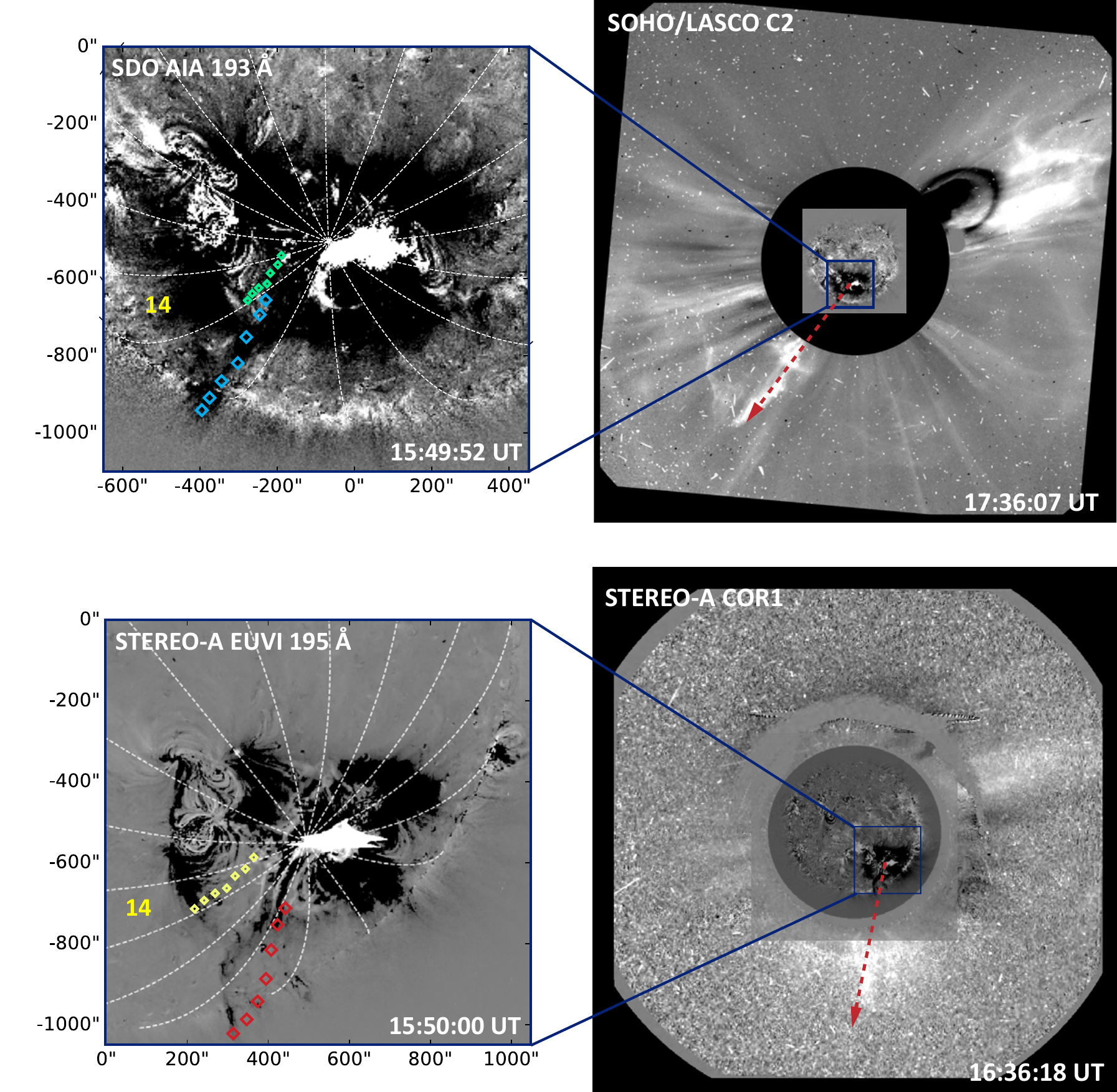}
\caption{Connection of the low coronal signatures with the CME propagation further out. Left panels: SDO AIA $193~\angstrom$ and STEREO-A EUVI $195~\angstrom$ base-difference images with reconstructed matching points of the eruptive filament (blue and red, respectively) 
and their orthogonal projections on the surface (green and yellow). The sector with the largest area of dimming is marked by its number (no.14). Right panels: the propagation of the CME seen from SOHO/LASCO C2 (top) and STEREO-A COR2 (bottom) coronagraphs at later time steps. Red arrows indicate the filament development as the bright substructure of the CME. Coronagraphic observations were obtained using JHelioviewer \citep{JHelioviewer}.}
\label{fig:orthogonal}
\end{figure}

Figure~\ref{fig:orthogonal} represents the connection of the low coronal signatures (left panels) with the CME propagation further out (right panels). The left panels show snapshots of SDO/AIA $193~\angstrom$ (top, 15:49:52~UT) and STEREO-A/EUVI $195~\angstrom$ (bottom, 15:50:00~UT) with matching points of the eruptive filament (blue and red, respectively, the same points as in Figure~\ref{fig:3d_snapshots}b) and their orthogonal projections on the surface (green for AIA and yellow for EUVI).
The dimming sectors on the solar surface are shown by a white grid. Solar EUV images show only the filament projections along the LOS of the spacecraft, and, as can be seen from Figure~\ref{fig:orthogonal} (left panels), the LOS and the orthogonal (which is independent of the viewpoint) projections of the filament are located in different sectors. At the same time, the distance between the LOS and orthogonal projections is significantly larger for STEREO-A (bottom), than for SDO (top), as STEREO-A reveals the location of the event in the southern-western hemisphere ($37.6^{\circ}$ east of Earth), whereas SDO gives a view of the event in the southern hemisphere close to the disk center. Note, that the orthogonal projections of the filament are located in the sector with the largest area of dimming (no.14).

The right panels in Figure~\ref{fig:orthogonal} show later observations of the CME propagation from SOHO/LASCO C2 (top) and STEREO-A COR2 (bottom) coronagraphs. This visual representation shows that the filament trajectory appears to look differently from these two viewpoints. The filament propagates almost as a straight line for SDO view. At the same time, STEREO observations reveal that a part of the filament structure decelerates and falls back to the solar surface.

\begin{figure}
\centering
\includegraphics[width=1\columnwidth]{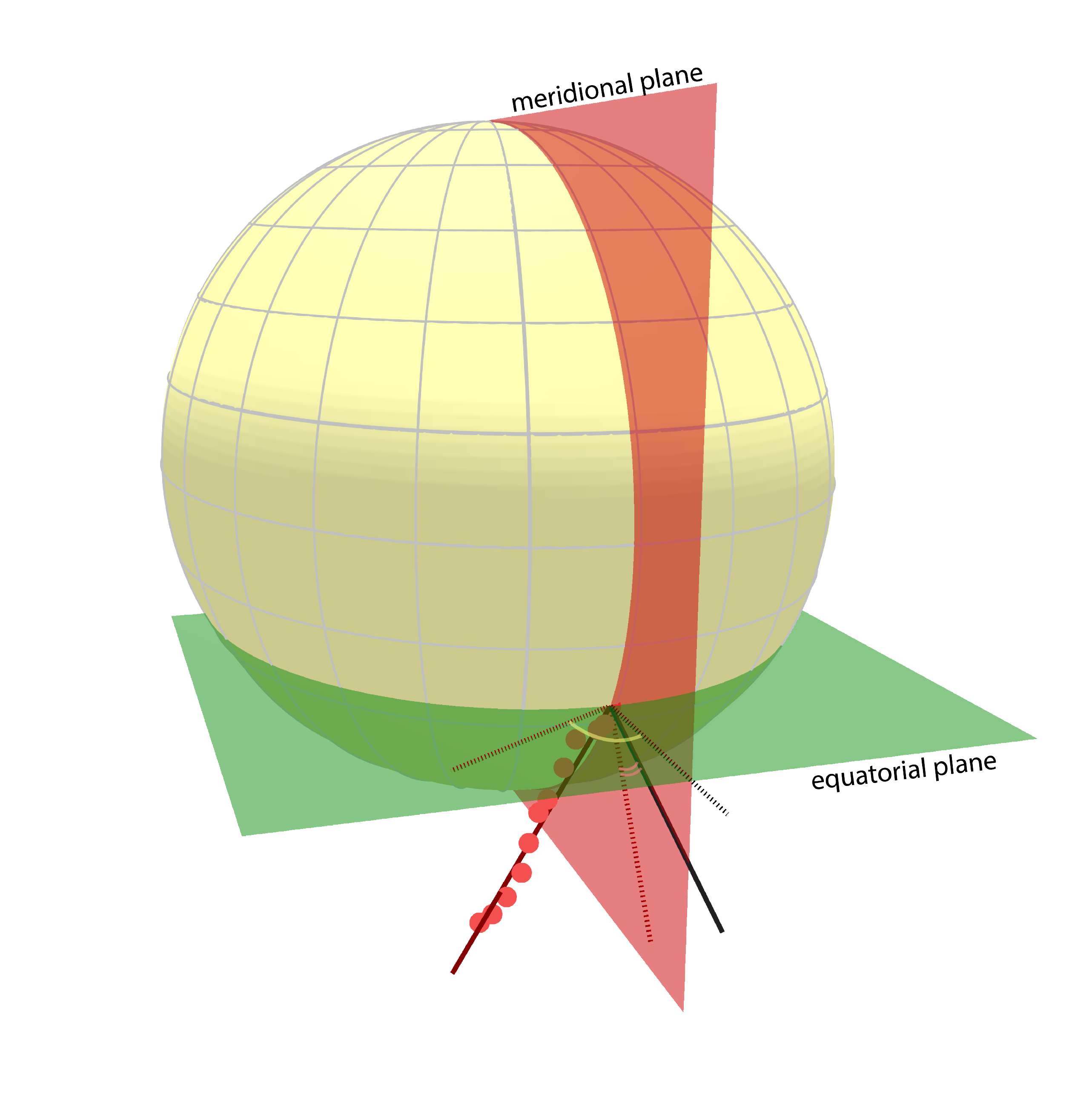}
\caption{3D-model of the Sun with the reconstructed filament. Red marks points show the reconstructed heights of the filament. The dark red line shows a linear fit, the red dashed lines represent projections onto equatorial (green) and meridional (red) planes. The black line indicates the radial direction, and the dashed black line - its projection onto the equatorial plane. Equatorial and meridional planes both go through the intersection of the linear fit with the solar surface (filament origin). The radial direction is located in  the meridional plane by definition. For clarity, the heliocentric coordinates grid is shown in gray. The inclination angle from the radial in the equatorial plane is shown by a yellow arc, and the inclination angle in the meridional plane - by two red arcs, respectively.}
\label{fig:filament}
\end{figure}

To gain a deeper understanding of the filament trajectory, we calculate its deflection from the radial direction in 3D space. To do so, we define a linear fit to the filament points constrained to the first reconstructed point with the lowest height. The angle between the linear fit and the radial direction is $54^{\circ}$. SDO/STEREO observations allowed us to reconstruct only one ``leg'' of the filament (the Eastern one), whereas the coronagraphic observations indicate another filament part, appearing earlier. \cite{sunasstar2022} also revealed this double structure of the eruption and estimated the velocities of the other part. We additionally analyzed the deflection of the filament propagation from radial by splitting the angle of deflection into two components. For that, we projected the linear fit and radial direction onto equatorial and meridional planes and calculated the 2D angles between projections in each plane. Both vectors (dark red and black lines, respectively) with their projections (dashed lines) are presented in Figure \ref{fig:filament}. We derive that the linear fit to the filament points is declined from radial by $64^{\circ}$ (yellow arc) to the eastern and $38^{\circ}$ (two red arcs) to the southern direction.

\subsubsection{Kinematics of erupting filament}
\label{sec:filament_kinematics}
\begin{figure}
\centering
\includegraphics[width=0.96\columnwidth]{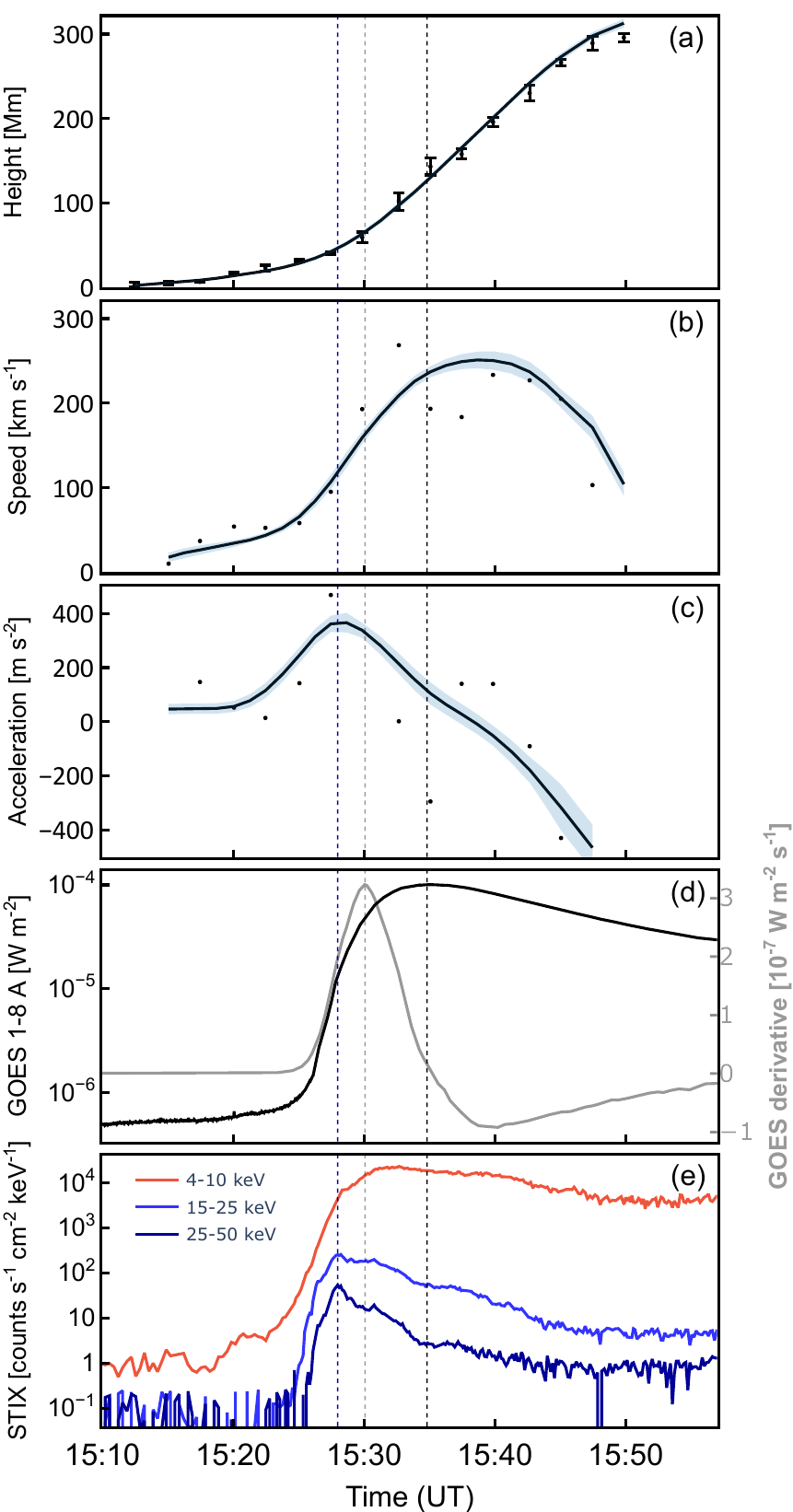}
\caption{Kinematics of the erupting filament and associated SXR flare evolution. 
(a) Height estimations of the filament (black markers). The error bar presents the 5-pixel shift in finding the matching point along the epipolar line. The corresponding black solid line indicates the smoothed height–time profile.
(b) Speed and (c) acceleration of the filament obtained by numerical differentiation of height-time data (dots) and smoothed profiles (lines). The shaded areas give the error ranges.
 (d) GOES $1-8~\angstrom$ SXR light curve (black curve, left Y-axis) and the corresponding change rate (gray curve, right Y-axis). (e) STIX count rates at 4–10~keV (red), 15–25~keV (blue), and 25–50~keV (dark blue) energies. The vertical dashed lines mark the peaks in the 25–50~keV STIX curve (dark blue), the GOES SXR flux (black), and its derivative (gray).}
\label{fig:heights}
\end{figure}

From the 3D reconstructions of the upper tip of the filament identified in each frame, we estimate its height and speed. Figure~\ref{fig:heights} shows the time evolution of the height (a), speed (b) and acceleration (c) of the filament in comparison to GOES soft X-rays (SXR) flux evolution of the associated flare and its change rate (d) and the STIX hard X-ray count rates at different energies (e).  
 In the panel (a) we show the resulting filament heights with black dots, while the error bar represents a $\pm$5-pixel shift in the AIA images in finding the matching point along the epipolar line. To obtain the filament velocity and acceleration profiles, we first smooth the filament heights data and obtain the first and second direct numerical derivatives. The smoothing technique, presented in \cite{podladchikova2017sunspot} extended to non-equidistant data, optimizes between two criteria in order to find a balance between data fidelity and smoothness of the approximating curve \citep[see also the applications in][]{veronig2018genesis, dissauer2019statistics, 2020ApJ...897L..36G, 2023A&A...672A..23S}. Using the derived acceleration profile, we then interpolate to equidistant data points (solid line in (c) panel) based on the minimization of the second derivatives and reconstruct the corresponding velocity (solid line in (b) panel) and height (solid line in the (a) panel) profiles by integration. We also obtain the errors of kinematic profiles by representing the reconstructed filament height, velocity and acceleration as an explicit the function of original errors of filament height–time data (blue shaded areas). The dots in panel (b) and (c) give the first and second time derivatives obtained directly from the height measurements.
As the eruption evolves, starting around 15:40~UT the upper filament parts become fainter. Thus, at later stages of observation, we may not be able to accurately identify the highest points of the filament, but instead, only the lower ones located further inside it. Consequently, the measured heights and velocities at these times may only be underestimated.
 In panel (d), we calculate the change rate of the GOES SXR flux as the first-order numerical derivative and smooth by a forward-backward exponential smoothing method \citep{Brown1963}. 
As can be seen from Figure~\ref{fig:heights}, the height of the filament increases up to 312$\pm$5~Mm within 37 minutes with a maximum speed of $\approx$250~\kmps. The filament shows a slow rise from 15:12 until 15:25~UT, then rapidly accelerates to $\approx$366~\mpss, reaching 52 Mm at 15:28~UT with a further gradual increase of speed. The peak of acceleration is cotemporal with peaks in the high energy STIX curve and a rapid increase of GOES flux. The overall evolution of the filament speed profile is preceding the associated flare flux: the flare started at $\approx$15:17~UT, reached its maximum around 15:35~UT, and went into a decline phase afterward.

\subsection{GCS-reconstruction of the CME}\label{sec:GCS}
\begin{figure}
\centering
\includegraphics[width=1\columnwidth]{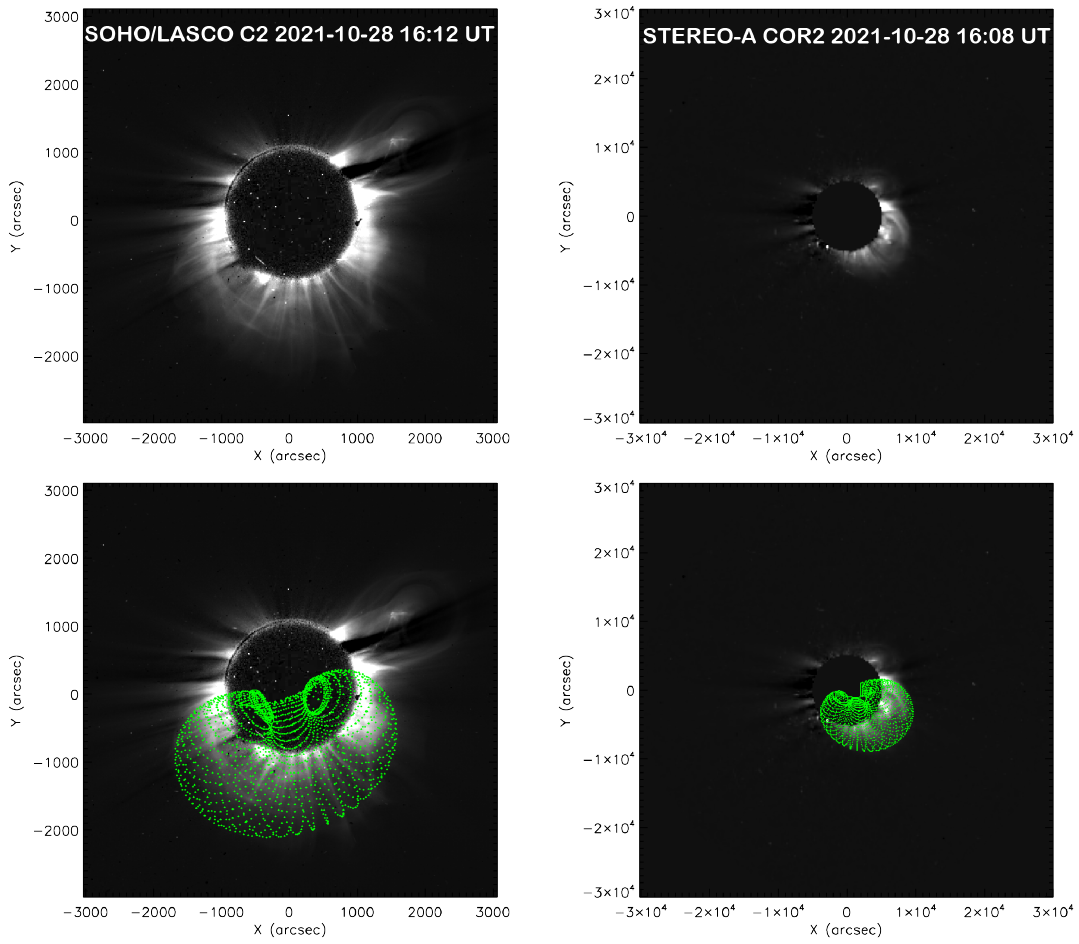}
\caption{Observations of the CME by SOHO/LASCO C2 (left) and STEREO-A COR2 (right) coronagraphs with the GCS-reconstructions (green mesh).}
\label{fig:GCS}
\end{figure}

We also analyzed the associated 3D CME structure and direction using the GCS model applied to the SOHO/LASCO C2 and STEREO-A/COR2 coronagraph data which have a FOV out to 6\Rsun~and 15\Rsun, respectively. The top panels in Figure~\ref{fig:GCS} show the observations of the CME by STEREO-A COR2 (left) and SOHO/LASCO C2 (right) coronagraphs. The bottom panels show the best fit (green mesh) resulting from the GCS 3D flux rope model when requiring that the boundary of the GCS model flux rope match the outer edge of the CME shape in COR2 (left) and LASCO (right) white-light images.

We obtain the following parameters of this reconstruction: heliocentric longitude $lon=357^{\circ}$, heliocentric latitude $lat=-30^{\circ}$, the height corresponding to the apex of the croissant $h=6$\Rsun, the tilt of the croissant axis to the solar equatorial plane $tilt=15^{\circ}$, half-angle measured between the apex and the central axis of a leg of the croissant $\alpha/2=60^{\circ}$, and aspect ratio $r=0.5$. The footprint positions of the croissant are located in [299\textdegree,$-46$\textdegree] and [55\textdegree,$-14$\textdegree].  As can be seen from Fig.\ref{fig:GCS}, the CME according to GCS reconstructions propagates almost radially in the southeastern direction.

 \subsection{Relationship between dimming, filament and CME directivity} 
 \label{sec:direction}

\begin{figure}
\centering
\includegraphics[width=1\columnwidth]{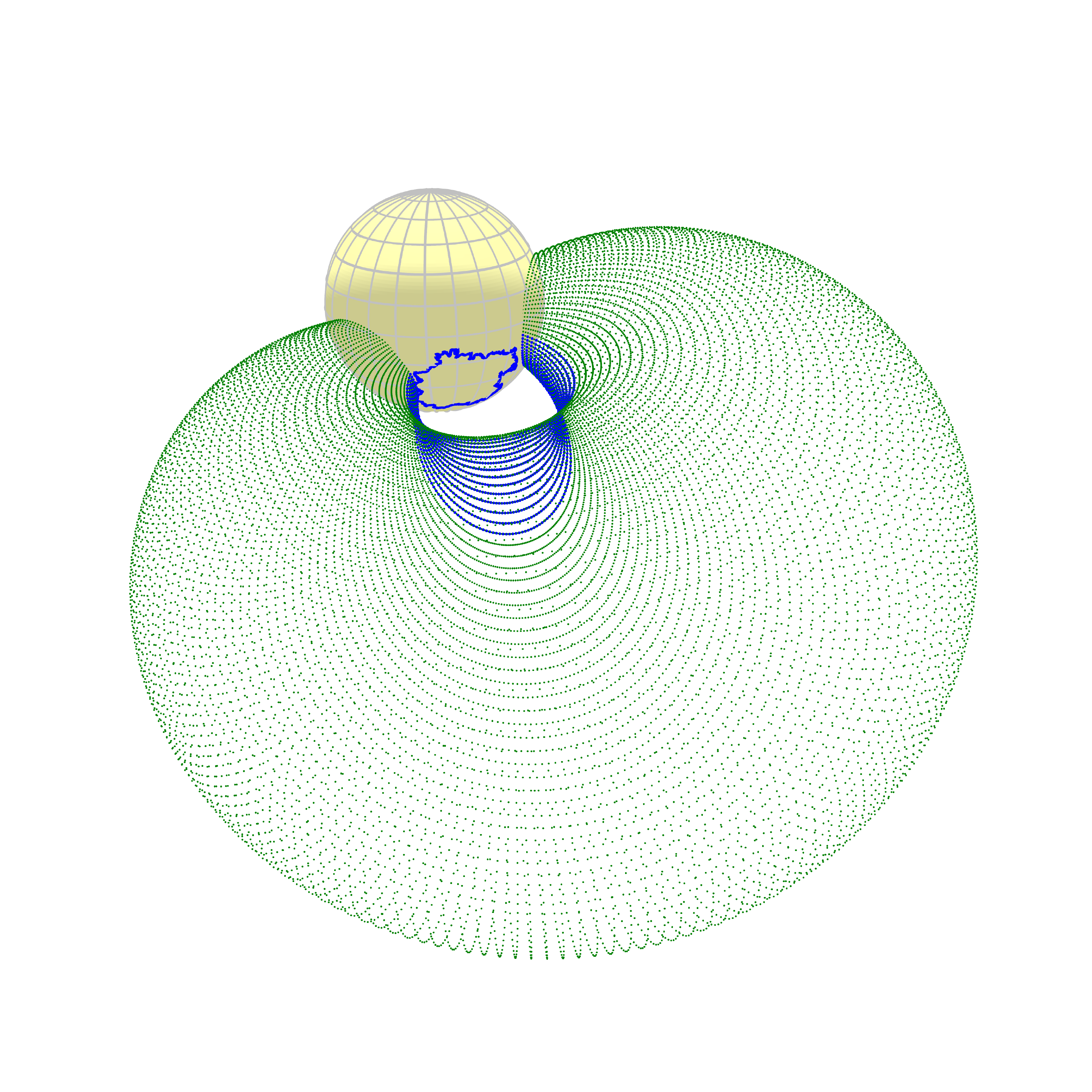}
\caption{3D-model of the Sun with the GCS croissant (green mesh with blue inner part) and coronal dimming (blue contour).}
\label{fig:GCS&dimming}
\end{figure}

Using the Python code provided in \cite{johan_l_freiherr_von_forstner_2021_5084818} we plotted the GCS structure on the 3D model of the Sun to compare it with the coronal dimming. Figure~\ref{fig:GCS&dimming} reveals a match between the total dimming region (blue contour) and the reconstructed shape and direction of the CME, represented by the GCS croissant (green mesh). Notably, the dimming contour precisely aligns with the inner portion (blue mesh) located between the two footpoints of the GCS croissant.  Our tracking of the dimming region extends from 15:00~UT up to 15:57~UT, while the CME structure was reconstructed for 16:12~UT. Assuming the CME propagates without deflection during this time span, our observations strongly suggest a spatial association between these features. To investigate this association more thoroughly, we compare both structures on the solar surface, using the orthogonal projection of the inner part (the closest one to the solar surface) and the location of GCS's primary axis. 

\begin{figure}
\centering
\includegraphics[width=1\columnwidth]{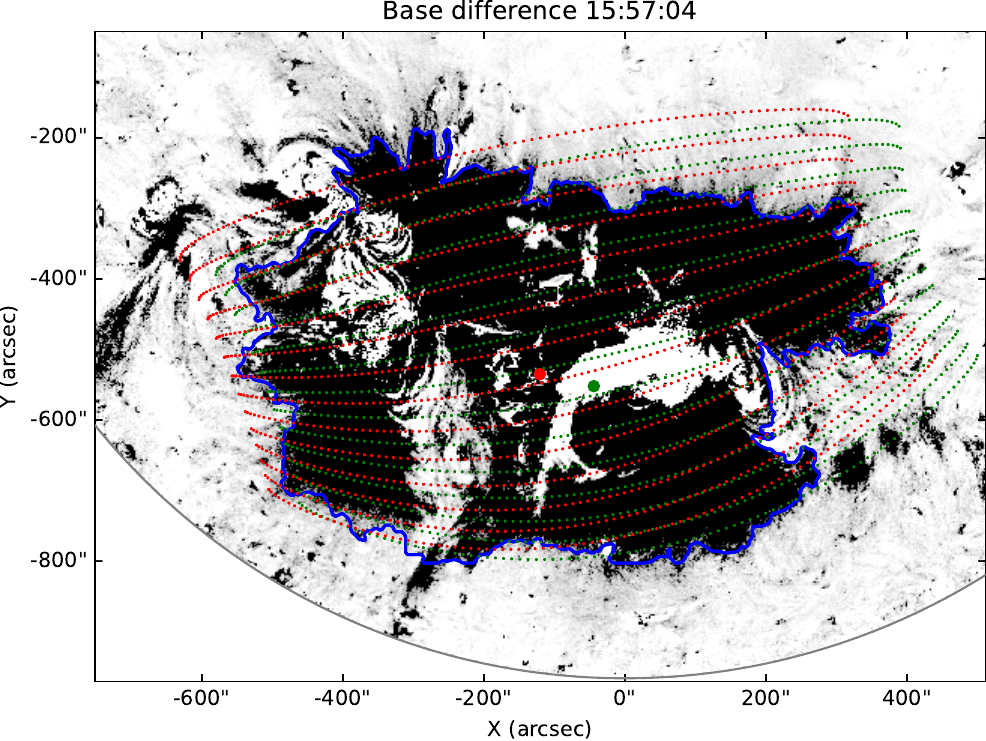}
\caption{AIA~193~\angstrom~base difference image showing the dimming region (blue contour) with an overlaid orthogonal projection of the inner part of GCS bubble (green mesh). The green marker indicates the source location derived from the GCS reconstruction, and the red marker represents the center of mass of the dimming region. The red mesh illustrates the orthogonal projection of the inner part of the GCS reconstruction, with the source point shifted to the center of mass location.} 
\label{fig:cm_inner_part}
\end{figure}

Figure~\ref{fig:cm_inner_part} shows AIA 193~\AA~image with the total cumulative dimming area (blue contour) and the projected inner part of the GCS reconstruction (green mesh). We mark the source point of the primary axis of the GCS croissant with a green marker. As can be seen, the projections (green) are mainly concentrated in the center of the dimming which allows us to link the 2D dimming with the 3d CME bubble. To explore this further, we also determine the center of mass of the dimming (red marker in Figure~\ref{fig:cm_inner_part}). For each pixel, we consider its position vector $\mathbf{r}_i = (x_i, y_i)$ and associate it with a weight $a_i$, where $i = 1, 2, \ldots, n$. The center of mass, denoted by $\mathbf{r}_{\text{cm}}$, is then computed using the formula:

\begin{align}
    \mathbf{r}_{\text{cm}} = \frac{\sum_{i=1}^{n} a_i \mathbf{r}_i}{\sum_{i=1}^{n} a_i}
\end{align}

Here, the weight of each pixel corresponds to its area on the solar surface. Consequently, the larger the area, the more significant its contribution to the overall dimming morphology. By analyzing the coordinates of the center of mass, we gain insights into the dimming's propagation relative to the eruption source (flare coordinate) and can compare it to the primary axis footpoint of the GCS croissant. The center of mass for the dimming has a heliographic longitude of $lon_{cm}=-8^{\circ}$ and a heliographic latitude of $lat_{cm}=-28.8^{\circ}$ and lies within the sector associated with the dominant dimming propagation. The center of mass is located $5^{\circ}$ to the West and $1.2^{\circ}$ to the North from the location of primary axis of GCS, which is within the uncertainty range of the reconstruction.

\begin{figure}
\centering
\includegraphics[width=1\columnwidth]{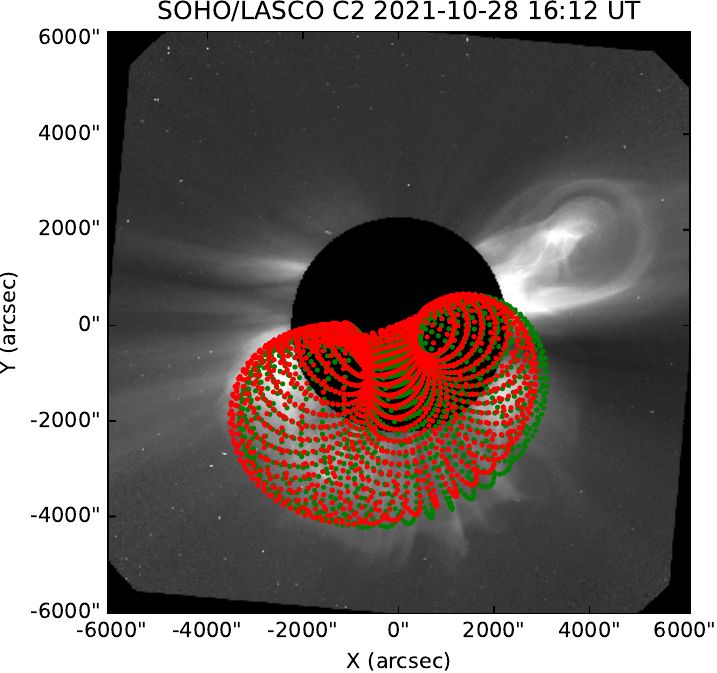}
\caption{
Comparison of the two versions of the GCS reconstruction on the SOHO/LASCO C2 16:12 UT image of the CME. Green mesh represents the GCS reconstruction from coronagraph observations, and red mesh - the same croissant with a changed location of source chosen by the dimming center of mass.
}
\label{fig:2GCS}
\end{figure}

We further create a modified GCS croissant with the same parameters but with a shifted primary axis to align with the center of mass of the dimming area. The orthogonal projection of this adjusted croissant is illustrated in Figure~\ref{fig:cm_inner_part} as red mesh. 
The full GCS croissant (green) with its shifted version (red) almost align in the SOHO/LASCO 16:12 UT coronagraph image (Figure~\ref{fig:2GCS}). Consequently, the center of mass can serve as an additional validation measure for obtaining the CME direction.

\begin{figure*}
\centering
\includegraphics[width=0.8\textwidth]{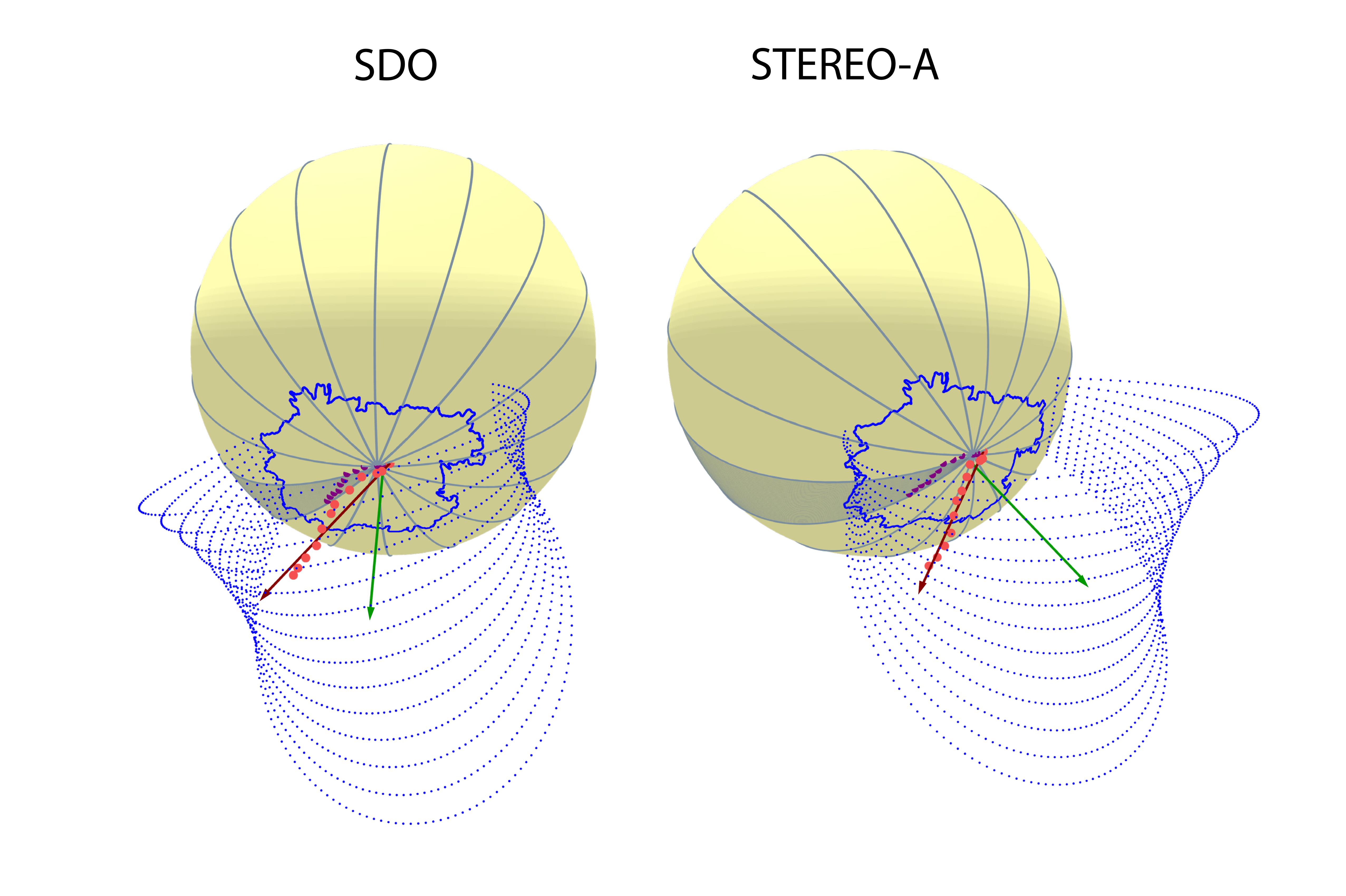}
\caption{3D-model of the Sun observed from SDO (left) and STEREO-A (right). The dimming borders are shown by blue contour, the sectors are drawn as light blue lines (the filled area indicates the sector of dominant dimming propagation). Red points mark the reconstructed filament heights, and purple points are their orthogonal projections on the surface. The inner part of the GCS reconstruction is plotted as blue mesh. The dark red line presents the straight line fit to the reconstructed filament positions. The green line shows the axis of the CME, represented by the GCS croissant. }
\label{fig:3d}
\end{figure*}

Figure~\ref{fig:3d} combines all results of our event analysis in one visualization of the studied eruptive features, shown from SDO and STEREO-A perspectives: the dimming region (blue contour) with the dominant dimming direction (filled sector), the reconstructed 3D filament motion (red points) with their orthogonal projections (purple points), and the modeled CME according to GCS reconstructions (inner part of the GCS bubble is shown as blue mesh, main axis - green vector). The dark red line shows a linear fit the reconstructed 3D filament points constrained to the first reconstructed point with the lowest height. 

With this visual representation, for the first time we can directly compare and relate the dimming evolution, the filament eruption, and the CME propagation with each other:

\begin{itemize}

    \item  The orthogonal projections of the erupting filament, which reflect the projected motion of this structure onto the solar surface, are located in the same sector as the dominant direction of the dimming evolution. Orthogonal projections are independent of the observation point, allowing us to establish a link between the 3D direction vector and the evolution of the dimming area on the solar surface. Consequently, the morphology of the coronal dimming serves as a reliable indicator for the lower-lying signatures associated with the eruption of the flux rope.

    \item  Despite the prevailing southwestern direction of the dimming propagation, it is important to consider the comprehensive dimming morphology, which also exhibits its widespread evolution around the eruption source. A compelling correspondence can be observed between the total dimming region and the reconstructed shape and direction of the CME. Specifically, the dimming contour reflects the inner part of the GCS croissant, indicating their spatial association.

    \item  The angle between the fit to the reconstructed filament height evolution and the apex of the GCS croissant is $50^{\circ}$. This value roughly matches with the CME half-width parameter of the GCS reconstruction ($\alpha/2=60^{\circ}$), which represents the angle between the apex and the leg of the CME croissant. This similarity suggests that the reconstructed filament can be reasonably associated with the Eastern leg of the GCS croissant, indicating a notable correspondence between the two structures.
    
\end{itemize}

\section{Discussion and Conclusions} \label{sec:support}

In this paper, we analyzed in detail the different features involved in the eruption of the 28 October 2021 X1.0 event, namely the erupting filament, the CME, and the coronal dimming evolution. 

Our research involved the reconstruction of both the low-lying filament eruption and the CME structure at a height of 6\Rsun~in order to determine the direction of the initial flux rope propagation in 3D space. Simultaneously, we defined the dominant propagation of the dimming growth based on the assessment of the evolution of the dimming area and the sector analysis on the solar surface. To compare the directions we calculated the orthogonal projections of the reconstructed points (both filament and GCS), which enabled us to track their motion projected onto the solar surface. 

We developed a method to estimate the dominant direction of the dimming development based on the assessment of the evolution of the dimming area using sector analysis. 
Initially, the dimming expansion spreads symmetrically around the source region, and subsequently, it undergoes further development towards the southeastern direction. The sector analysis of the cumulative dimming area clearly indicates that the dominant direction of total dimming growth aligns with one of the southeastern sectors.

As we consider the dimming region on the solar sphere, we have introduced a novel technique that involves identifying the surface area for each pixel. This approach effectively minimizes projection effects, as it takes into account the spherical projection of each individual pixel of the dimming area. We have made the code for the calculation of the correction publicly available on GitHub.
    
A filament was observed both by SDO and STEREO-A up to $\approx$312~Mm in height, which allowed us to reconstruct its evolution in 3D. The erupting filament starts to rise at a slow speed at $\approx$15:12~UT, then rapidly accelerates between $\approx$15:25--15:30 UT which  coincides with the flare rise phase in the GOES 1--8~\angstrom~SXR flux, while the filament impulsive acceleration occurs co-temporal with the STIX high-energy hard X-ray emission at 15–25~keV and 25–50~keV (see Fig. \ref{fig:heights}). The maximum speed and acceleration of the erupting filament area $\approx$250~\kmps~and $\approx$366~\mpss, respectively. A similar filament evolution was also reported in \cite{devi2022extreme}, where the time-distance plots of the filament heights in 2D were used to derive the height and kinematics profiles. These authors also report a non-linear acceleration at the same range, which they relate to the presence of torus instability in the eruption behavior. 
    
The reconstructed filament heights projected to the solar surface coincide with the dominant direction of the dimming expansion. This allows us for the first time to provide a direct comparison of the dimming and filament directions within a unified coordinate system.
We find that the dimming accurately reflects the part where the filament eruption occurred low in the corona.

The significant inclination of the 3D filament motion from radial direction ($64^{\circ}$ to the East and $38^{\circ}$ to the South) reveals a double structure of the flux rope. Both filament parts are observed as the most bright and most dense CME material in the coronagraph images (Figure~\ref{fig:orthogonal}, right panels).
By modeling the propagation of the flux rope using GCS reconstruction, we find that the CME primarily propagates radially in the southeastern direction. We obtain a deviation of $50^{\circ}$ between the direction of the CME and the erupting filament. This indicates that the portion of the filament used for the 3D reconstruction, which is clearly observed, does not correspond to the apex of the CME flux rope. Instead, it is associated with one of the legs of the flux rope. In addition, the close agreement between the deviation angle value and the half-width parameter of the GCS reconstruction ($\alpha/2=60^{\circ}$) further supports the correspondence between the reconstructed filament and the eastern leg of the GCS croissant.

The strongly non-radial behavior of the erupting filament obtained from the 3D reconstruction could also indicate a pronounced lateral expansion of the flux rope, which contrasts with the assumption of self-similar propagation in the GCS reconstruction of the CME. This would lead to non-radial propagation of the CME leg, even if its apex is propagating radially. \cite{prominence} also discussed whether the evolution of the filament in the low corona can be used as a predictor of CME direction, as observed in coronagraph data.

We observed distinct signatures of both components of the flux rope in the evolution of the dimming region. While the dominant propagation of the dimming follows a southwestern direction, reflecting the filament evolution, it is important to acknowledge the comprehensive morphology of the dimming, which shows an extensive evolution around the eruption source. Our analysis reveals a compelling relation between the total dimming region and the reconstructed shape and direction of the CME. Specifically, the dimming contour closely aligns with the inner part of the GCS croissant, while the center mass of the dimming region is situated in close proximity to the projection of the primary axis of the GCS reconstruction, indicating a clear spatial association between dimming and CME.

At the same time, for our case study, the direction of the filament eruption is not related directly to the
direction of the CME expansion due to a significant deviation angle between the reconstructed filament heights and the CME (GCS) axis. The filament material is lying in the lower portion of the expanding flux rope \citep{2014ApJ...780...28C}. With that, we point out that the filament itself cannot always be used as a reliable proxy to predict the main direction of CME propagation.

These findings highlight that the dimmings evolution reflects both the 
direction of a low-lying erupting magnetic structure (filament) and  the global propagation of CMEs.
Further research efforts need to consider both the dominant dimming direction and the dimming morphology to provide a link from 2D dimming information to the initial 3D direction of the CME/flux rope, enhancing our understanding of its early dynamics.

\section{Acknowledgements} 
\begin{acknowledgements}   
G.C. and T.P. acknowledge support from the Russian Science Foundation under the project 23-22-00242. K.D. acknowledges support from NASA under award No. 80NSSC21K0738 and the NSF under AGS-ST Grant 2154653. A.M.V and E.C.M.D. acknowledge the Austria Science Fund (FWF): project no. 14555-N. M.D. acknowledges the support by the Croatian Science Foundation under the project IP-2020-02-9893 (ICOHOSS). SDO data are courtesy of the NASA/SDO AIA and HMI science teams. The SOHO/LASCO data used here are produced by a consortium of the Naval Research Laboratory (USA), Max-Planck-Institut f\"ur Aeronomie (Germany), Laboratoire d'Astronomie (France), and the University of Birmingham (UK). The STEREO/SECCHI data are produced by an international consortium of the Naval Research Laboratory (USA), Lockheed Martin Solar and Astrophysics Lab (USA), NASA Goddard Space Flight Center (USA), Rutherford Appleton Laboratory (UK), University of Birmingham (UK), Max-Planck-Institut f\"ur Sonnenforschung (Germany), Centre Spatiale de Li\`ege (Belgium), Institut d'Optique Th\'eorique et Appliqu\'ee (France), and Institut d'Astrophysique Spatiale (France). Solar Orbiter is a mission of international cooperation between ESA and NASA, operated by ESA. We thank the referee for valuable comments on this study.
\end{acknowledgements}

\bibliography{sample631}{}
\bibliographystyle{aa}

\appendix 
\section{Pixel area calculation} \label{sec:appendix}
The method to define the main direction of the dimming propagation, described in this work, strongly relies on the dimming area parameter. The images of the Sun provide only a 2D representation of the actual 3D solar surface. Projection effects lead to the different weights of each pixel, e.g. closer to the limb the pixel area is larger than one closer to the Sun center. Thus, for an accurate definition of the area, we present a method to identify the surface area of a sphere for each pixel.

The equation of the sphere is given by 
\begin{equation}
x^2+y^2+z^2=R^2
\end{equation}
Here, $R$ is the radius of the sphere in kilometers, and $(x,y,z)$ are the coordinates of points of a sphere. The surface of a sphere can be described as a function of two variables
\begin{equation}\label{Eq_z}
z=f(x,y)=\sqrt{R^2-x^2-y^2}
\end{equation}

The projection of the sphere onto the $XY$ plane is a circle belonging to the $XOY$ plane, where $O$ is the center of the sphere. The surface area of a sphere that has the region $S$ as its projection onto the $XOY$ plane is given by \citep{Bronstein2005} 
\begin{equation}\label{Int_S}
     \sigma = \int_{S}\int \sqrt{1+ \left(\frac{\partial z}{\partial x}\right)^2 + \left(\frac{\partial z}{\partial y}\right)^2} dx dy
\end{equation}

As follows from Equation~(\ref{Eq_z})
\begin{equation}\label{Eq_der}
\begin{array}{cc}
     \frac{\partial z}{\partial x}=-\frac{x}{\sqrt{R^2-x^2-y^2}}\\
     \frac{\partial z}{\partial y}=-\frac{y}{\sqrt{R^2-x^2-y^2}}     
\end{array}
\end{equation}
Thus using Equation~(\ref{Eq_der}), Equation~(\ref{Int_S}) can be rewritten as
\begin{equation}\label{Int_S1}
     \sigma = R\int_{S}\int \frac{1}{\sqrt{R^2-x^2-y^2}} dx dy
\end{equation}

\begin{figure}
\centering
\includegraphics[width=1\columnwidth]{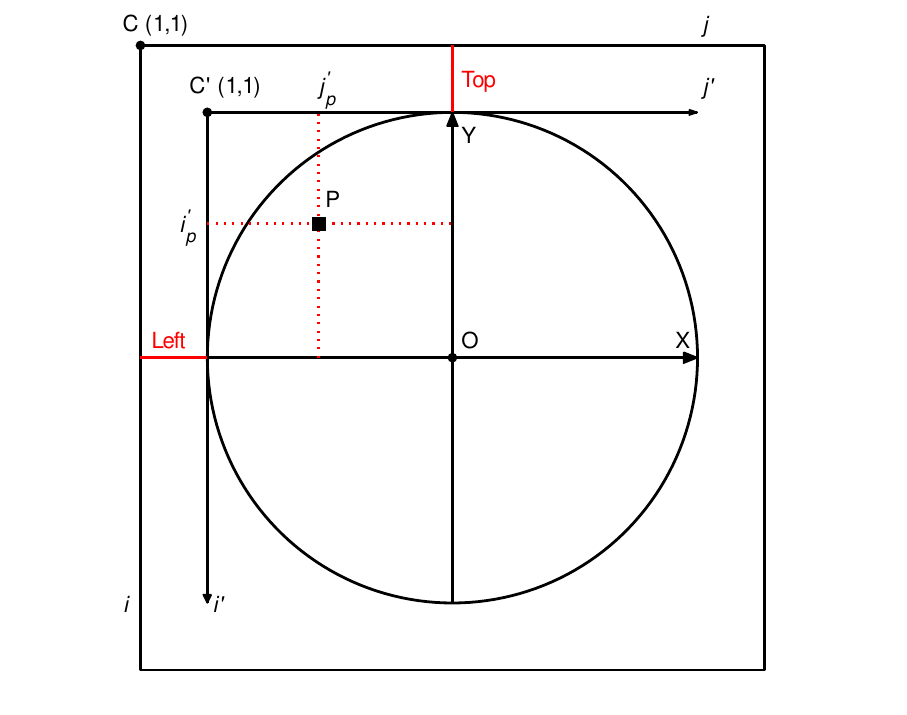}
\caption{Scheme of a solar image. The outer square shows the boundaries of a solar image from a spacecraft. The marker C at the top left corner indicates the pixel with the coordinates (1,1), $i$ - row number, $j$ - column number. The marker C' also shows the pixel with the coordinates (1,1), but in a new coordinate system with rows $i'$ and columns $j'$ connected to $i$ and $j$ via  Equation~(\ref{Eq_coord}). ``Top''/``Left'' show the distance (in pixels) from the top/left border of the image to the beginning of the Sun. $P$ is a pixel, and $(i_{p}^{'},j_{p}^{'})$ are the coordinates of its center.}
\label{fig:pixel}
\end{figure}

Further, we present how to estimate the surface area of a sphere, which projection $S$ onto the $XOY$ plane has a size of 1 pixel. Figure~\ref{fig:pixel} shows the scheme of a solar image, where the outer square shows the boundaries of a solar image from a spacecraft. The marker C at the top left corner indicates the pixel with the coordinates (1,1), $i$, and $j$ mark row and column numbers. The marker C' also shows the pixel with the coordinates (1,1), but in a new coordinate system with rows $i'$ and columns $j'$ derived from $i$ and $j$ in the following way 
\begin{equation}\label{Eq_coord}
\begin{array}{cc}
     i' = i-Top \\
     j' = j-Left     
\end{array}
\end{equation}
Here, the Top/Left show the distance (in pixels) from the top/left border of the image to the beginning of the Sun. $P$ is a pixel, which area we are estimating, and $(i_{p}^{'},j_{p}^{'})$ are the coordinates of its center. Pixel $P$ is a square with a side $a$, which size is defined as
\begin{equation}\label{Eq_a}
a=\frac{R}{R_{P}}
\end{equation}
Here, $R$ is the radius of the Sun in kilometers, $R_{P}$ - radius of the Sun in pixels. 

A coordinate $Y$ of the lower pixel boundary is defined by 
\begin{equation}\label{Eq_Y_lower}
a(R_{P}-i'-0.5)
\end{equation}
A coordinate $Y$ of the upper pixel boundary is given by 
\begin{equation}\label{Eq_Y_upper}
a(R_{P}-i'+0.5)
\end{equation}
A coordinate $X$ of the left pixel boundary is determined as 
\begin{equation}\label{Eq_X_left}
a(-R_{P}+j'-0.5)
\end{equation}
A coordinate $X$ of the right pixel boundary is defined by 
\begin{equation}\label{Eq_X_right}
a(-R_{P}+j'+0.5)
\end{equation}
Taking into account Equations~(\ref{Eq_Y_lower}--\ref{Eq_X_right}), and 
replacing the calculation of the double integral by the successive calculation of two simple integrals, Equation~(\ref{Int_S1}) can be rewritten as
\begin{equation}\label{Eq_two_integrals}
\sigma=R\int_{a(-R_{P}+j'-0.5)}^{a(-R_{P}+j'+0.5)} dx \int_{a(R_{P}-i'-0.5)}^{a(R_{P}-i'+0.5)} \frac{1}{\sqrt{R^2-x^2-y^2}}  dy
\end{equation}

Further, we solve the internal integral from Equation~(\ref{Eq_two_integrals}) analytically and the external one numerically. The internal integral from Equation~(\ref{Eq_two_integrals})  can be rewritten as
\begin{equation}\label{Eq_integral_internal}
\int_{a(R_{P}-i'-0.5)}^{a(R_{P}-i'+0.5)} \frac{1}{\sqrt{R^2-x^2}\sqrt{1-\frac{y^2}{R^2-x^2}}}  dy 
\end{equation}
Let us use the change of variables in Equation~(\ref{Eq_integral_internal})
\begin{equation}\label{Eq_change_var}
\begin{array}{cc}
     u=\frac{y}{\sqrt{R^2-x^2}} \\
     du=\frac{dy}{\sqrt{R^2-x^2}}
\end{array}
\end{equation}
Thus, Equation~(\ref{Eq_integral_internal}) can be rewritten as
\begin{equation}\label{Eq_integral_internal_change}
\int_{\frac{a(R_{P}-i'-0.5)}{\sqrt{R^2-x^2}}}^{\frac{a(R_{P}-i'+0.5)}{\sqrt{R^2-x^2}}} \frac{1}{\sqrt{1-u^2}}  du 
\end{equation}
The solution of integral represented by Equation~(\ref{Eq_integral_internal_change}) is given by
\begin{equation}\label{Eq_integral_internal_solution}
arcsin(\frac{a(R_{P}-i'+0.5)}{\sqrt{R^2-x^2}})-arcsin(\frac{a(R_{P}-i'-0.5)}{\sqrt{R^2-x^2}})
\end{equation}

Further, we solve the external integral of the resulted function given by Equation~(\ref{Eq_integral_internal_solution}) using the mean rectangle method \citep{Calculus2017}. 
First, we divide the integration interval by N regions 
\begin{equation}\label{Eq_a_N}
dx=a/N
\end{equation}
Here, $N=1000$. Then we calculate the integral by finding a sum of the areas of $N$ rectangles defined by
\begin{equation}\label{Eq_Rectangle}
\begin{array}{cc}
A_{k}=R(arcsin(\frac{a(R_{P}-i'+0.5)}{\sqrt{R^2-(x_{k}+\frac{dx}{2})^2}})-\\
-arcsin(\frac{a(R_{P}-i'-0.5)}{\sqrt{R^2-(x_{k}+\frac{dx}{2})^2}}))dx
\end{array}
\end{equation}
Here, $k=1,2,\cdots,N$; $A_{k}$ is calculated at the middle of each region, and $x(k)$ takes the values within integration limits from $a(-R_{P}+j'-0.5)$ to $a(-R_{P}+j'+0.5)$ and for each $k$ is estimated as
\begin{equation}\label{Eq_xk}
x(k)=a(-R_{P}+j-0.5)+(k-1)dx 
\end{equation}
The resulting area of a pixel is defined by the sum of $x_k$.

\begin{figure}
\centering
\includegraphics[width=1\columnwidth]{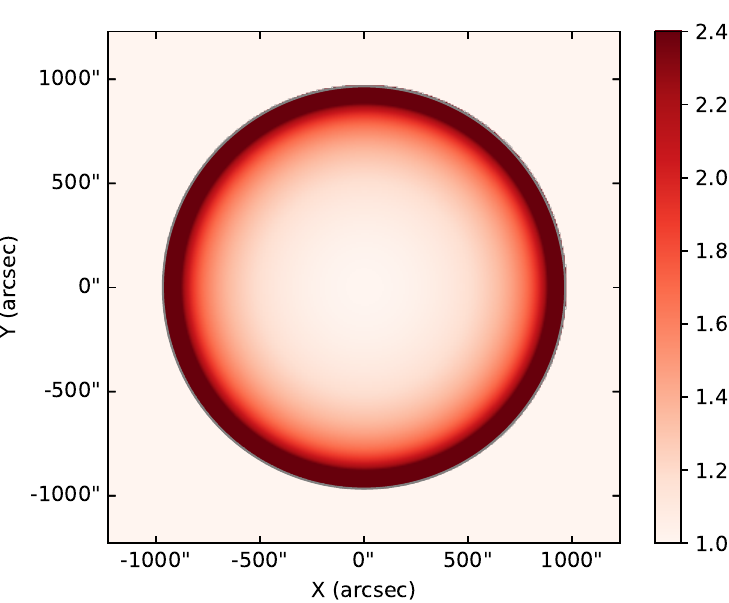}
\caption{Correction area map for the full solar disk at the currently observed radius of the Sun. The colorbar shows the ratio of every pixel area to the average pixel and increases closer to the limb.}
\label{fig:area_map}
\end{figure}

With this method, we calculate the area of every pixel of the solar disk and construct a correction map, where each pixel is weighted by its area (Figure \ref{fig:area_map}). The colorbar shows the ratio of each pixel area to the square pixel in the center of the solar disk. Any mask of the dimming region can be applied to this map to obtain the final dimming area - we already showed the corrected dimming mask for the 2021 October 28 event in Figure \ref{fig:dimming_bd_time}. 

\begin{figure}
\centering
\includegraphics[width=1\columnwidth]{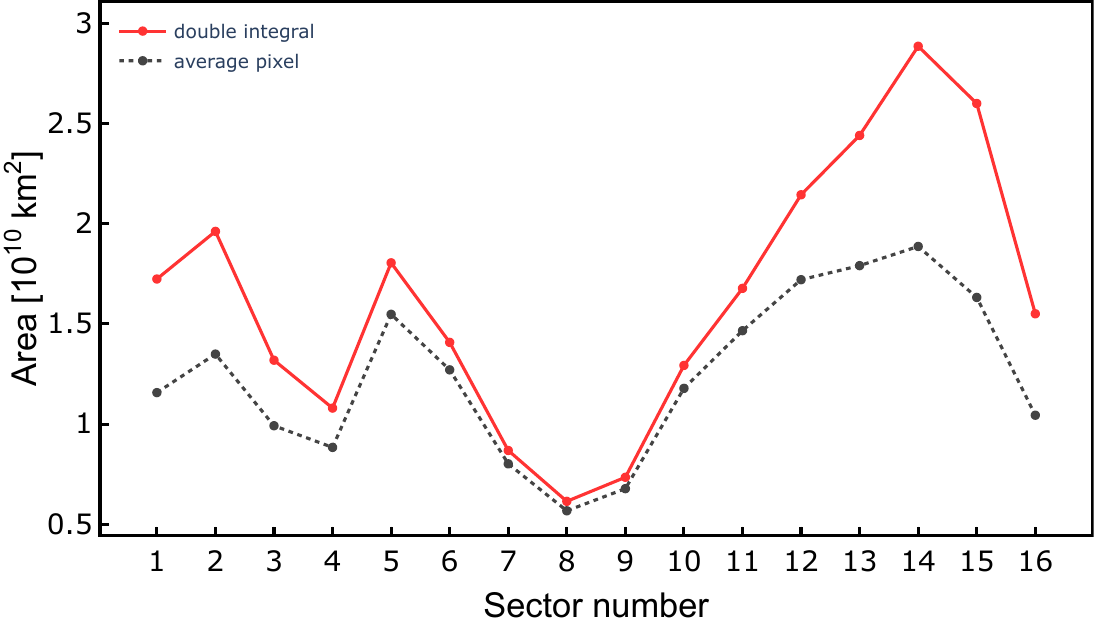}
\caption{Dimming area on the plane (dashed black curve) and over the surface (blue curve) calculated  for each sector.}
\label{fig:compare_area}
\end{figure}

Figure \ref{fig:compare_area} shows the difference between calculating the area from the number of pixels and considering their spherical representation. We can see that the suggested calibration creates significant changes in the total area values. Because of that reason, we use the derived spherical projections of the pixels to derive the area parameter. 

We propose adopting this approach for a precise correction of all on-disk pixels. Particularly it will be useful for the pixels located close to the limb (beyond $60^{\circ}$), disregarded in the classical correction method using a factor of $1/\cos\alpha$ \citep{hagenaar2001ephemeral}. By implementing this alternative method, we can achieve enhanced accuracy in the correction process for solar images.

\end{document}